\documentclass[prd,twocolumn,superscriptaddress,showpacs,amsmath,amssymb]{revtex4}
\usepackage{graphicx}
\usepackage{epsfig}
\usepackage{dcolumn}
\usepackage{bm}
\usepackage{slashed}
\usepackage{gensymb} 

\usepackage{multirow}
\usepackage{comment}
\excludecomment{printrobustnotes}
\excludecomment{printallnotes}

\usepackage{mathrsfs} 

\usepackage{url}
\usepackage{xspace}

\newcommand{\ttbar}{t\overline{t}}

\newcommand{\ppbar}{ {\it p} \bar{{\it p}} }
\newcommand{\qqbar}{q\bar{q}}

\newcommand{\met}{\slashed{E}}
\newcommand{\MTreco}{m_t^{reco}}

\newcommand{\MET}{{\not \!\!E_T}}

\setkeys{Gin}{width=.9\columnwidth, height=.75\textheight,
keepaspectratio}
\newenvironment{cfigure}[1][tbp]{\begin{figure}[#1]\centering}{\end{figure}}
\newenvironment{cfigure1c}[1][tbp]{\begin{figure*}[#1]\centering}{\end{figure*}}
\newenvironment{ctable1c}[1][tbp]{\begin{table*}[#1]\centering}{\end{table*}}

\newcommand{\note}[1]{\xspace}
\newcommand{\robustnote}[1]{\xspace}
\begin{printrobustnotes}
\renewcommand{\robustnote}[1]{\textbf{\textit{#1}}\xspace}
\end{printrobustnotes}
\begin{printallnotes}
\renewcommand{\note}[1]{\textbf{\textit{#1}}\xspace}
\renewcommand{\robustnote}[1]{\textbf{\textit{#1}}\xspace}
\end{printallnotes}
\newcommand{\px}     {\ensuremath{p_{x}}\xspace}
\newcommand{\py}     {\ensuremath{p_{y}}\xspace}

\begin{document}


\title{Measurement of the top quark mass at CDF using the ``neutrino $\phi$ weighting'' template method on a lepton plus isolated track sample}
\affiliation{Institute of Physics, Academia Sinica, Taipei, Taiwan 11529, Republic of China} 
\affiliation{Argonne National Laboratory, Argonne, Illinois 60439} 
\affiliation{University of Athens, 157 71 Athens, Greece} 
\affiliation{Institut de Fisica d'Altes Energies, Universitat Autonoma de Barcelona, E-08193, Bellaterra (Barcelona), Spain} 
\affiliation{Baylor University, Waco, Texas  76798} 
\affiliation{Istituto Nazionale di Fisica Nucleare Bologna, $^x$University of Bologna, I-40127 Bologna, Italy} 
\affiliation{Brandeis University, Waltham, Massachusetts 02254} 
\affiliation{University of California, Davis, Davis, California  95616} 
\affiliation{University of California, Los Angeles, Los Angeles, California  90024} 
\affiliation{University of California, San Diego, La Jolla, California  92093} 
\affiliation{University of California, Santa Barbara, Santa Barbara, California 93106} 
\affiliation{Instituto de Fisica de Cantabria, CSIC-University of Cantabria, 39005 Santander, Spain} 
\affiliation{Carnegie Mellon University, Pittsburgh, PA  15213} 
\affiliation{Enrico Fermi Institute, University of Chicago, Chicago, Illinois 60637}
\affiliation{Comenius University, 842 48 Bratislava, Slovakia; Institute of Experimental Physics, 040 01 Kosice, Slovakia} 
\affiliation{Joint Institute for Nuclear Research, RU-141980 Dubna, Russia} 
\affiliation{Duke University, Durham, North Carolina  27708} 
\affiliation{Fermi National Accelerator Laboratory, Batavia, Illinois 60510} 
\affiliation{University of Florida, Gainesville, Florida  32611} 
\affiliation{Laboratori Nazionali di Frascati, Istituto Nazionale di Fisica Nucleare, I-00044 Frascati, Italy} 
\affiliation{University of Geneva, CH-1211 Geneva 4, Switzerland} 
\affiliation{Glasgow University, Glasgow G12 8QQ, United Kingdom} 
\affiliation{Harvard University, Cambridge, Massachusetts 02138} 
\affiliation{Division of High Energy Physics, Department of Physics, University of Helsinki and Helsinki Institute of Physics, FIN-00014, Helsinki, Finland} 
\affiliation{University of Illinois, Urbana, Illinois 61801} 
\affiliation{The Johns Hopkins University, Baltimore, Maryland 21218} 
\affiliation{Institut f\"{u}r Experimentelle Kernphysik, Universit\"{a}t Karlsruhe, 76128 Karlsruhe, Germany} 
\affiliation{Center for High Energy Physics: Kyungpook National University, Daegu 702-701, Korea; Seoul National University, Seoul 151-742, Korea; Sungkyunkwan University, Suwon 440-746, Korea; Korea Institute of Science and Technology Information, Daejeon, 305-806, Korea; Chonnam National University, Gwangju, 500-757, Korea} 
\affiliation{Ernest Orlando Lawrence Berkeley National Laboratory, Berkeley, California 94720} 
\affiliation{University of Liverpool, Liverpool L69 7ZE, United Kingdom} 
\affiliation{University College London, London WC1E 6BT, United Kingdom} 
\affiliation{Centro de Investigaciones Energeticas Medioambientales y Tecnologicas, E-28040 Madrid, Spain} 
\affiliation{Massachusetts Institute of Technology, Cambridge, Massachusetts  02139} 
\affiliation{Institute of Particle Physics: McGill University, Montr\'{e}al, Qu\'{e}bec, Canada H3A~2T8; Simon Fraser University, Burnaby, British Columbia, Canada V5A~1S6; University of Toronto, Toronto, Ontario, Canada M5S~1A7; and TRIUMF, Vancouver, British Columbia, Canada V6T~2A3} 
\affiliation{University of Michigan, Ann Arbor, Michigan 48109} 
\affiliation{Michigan State University, East Lansing, Michigan  48824}
\affiliation{Institution for Theoretical and Experimental Physics, ITEP, Moscow 117259, Russia} 
\affiliation{University of New Mexico, Albuquerque, New Mexico 87131} 
\affiliation{Northwestern University, Evanston, Illinois  60208} 
\affiliation{The Ohio State University, Columbus, Ohio  43210} 
\affiliation{Okayama University, Okayama 700-8530, Japan} 
\affiliation{Osaka City University, Osaka 588, Japan} 
\affiliation{University of Oxford, Oxford OX1 3RH, United Kingdom} 
\affiliation{Istituto Nazionale di Fisica Nucleare, Sezione di Padova-Trento, $^y$University of Padova, I-35131 Padova, Italy} 
\affiliation{LPNHE, Universite Pierre et Marie Curie/IN2P3-CNRS, UMR7585, Paris, F-75252 France} 
\affiliation{University of Pennsylvania, Philadelphia, Pennsylvania 19104}
\affiliation{Istituto Nazionale di Fisica Nucleare Pisa, $^z$University of Pisa, $^{aa}$University of Siena and $^{bb}$Scuola Normale Superiore, I-56127 Pisa, Italy} 
\affiliation{University of Pittsburgh, Pittsburgh, Pennsylvania 15260} 
\affiliation{Purdue University, West Lafayette, Indiana 47907} 
\affiliation{University of Rochester, Rochester, New York 14627} 
\affiliation{The Rockefeller University, New York, New York 10021} 
\affiliation{Istituto Nazionale di Fisica Nucleare, Sezione di Roma 1, $^{cc}$Sapienza Universit\`{a} di Roma, I-00185 Roma, Italy} 

\affiliation{Rutgers University, Piscataway, New Jersey 08855} 
\affiliation{Texas A\&M University, College Station, Texas 77843} 
\affiliation{Istituto Nazionale di Fisica Nucleare Trieste/Udine, I-34100 Trieste, $^{dd}$University of Trieste/Udine, I-33100 Udine, Italy} 
\affiliation{University of Tsukuba, Tsukuba, Ibaraki 305, Japan} 
\affiliation{Tufts University, Medford, Massachusetts 02155} 
\affiliation{Waseda University, Tokyo 169, Japan} 
\affiliation{Wayne State University, Detroit, Michigan  48201} 
\affiliation{University of Wisconsin, Madison, Wisconsin 53706} 
\affiliation{Yale University, New Haven, Connecticut 06520} 
\author{T.~Aaltonen}
\affiliation{Division of High Energy Physics, Department of Physics, University of Helsinki and Helsinki Institute of Physics, FIN-00014, Helsinki, Finland}
\author{J.~Adelman}
\affiliation{Enrico Fermi Institute, University of Chicago, Chicago, Illinois 60637}
\author{T.~Akimoto}
\affiliation{University of Tsukuba, Tsukuba, Ibaraki 305, Japan}
\author{B.~\'{A}lvarez~Gonz\'{a}lez$^s$}
\affiliation{Instituto de Fisica de Cantabria, CSIC-University of Cantabria, 39005 Santander, Spain}
\author{S.~Amerio$^y$}
\affiliation{Istituto Nazionale di Fisica Nucleare, Sezione di Padova-Trento, $^y$University of Padova, I-35131 Padova, Italy} 

\author{D.~Amidei}
\affiliation{University of Michigan, Ann Arbor, Michigan 48109}
\author{A.~Anastassov}
\affiliation{Northwestern University, Evanston, Illinois  60208}
\author{A.~Annovi}
\affiliation{Laboratori Nazionali di Frascati, Istituto Nazionale di Fisica Nucleare, I-00044 Frascati, Italy}
\author{J.~Antos}
\affiliation{Comenius University, 842 48 Bratislava, Slovakia; Institute of Experimental Physics, 040 01 Kosice, Slovakia}
\author{G.~Apollinari}
\affiliation{Fermi National Accelerator Laboratory, Batavia, Illinois 60510}
% ...

\collaboration{CDF Collaboration\footnote{With visitors from $^a$University of Massachusetts Amherst, Amherst, Massachusetts 01003,
$^b$Universiteit Antwerpen, B-2610 Antwerp, Belgium, 
$^c$University of Bristol, Bristol BS8 1TL, United Kingdom,
$^d$Chinese Academy of Sciences, Beijing 100864, China, 
$^e$Istituto Nazionale di Fisica Nucleare, Sezione di Cagliari, 09042 Monserrato (Cagliari), Italy,
$^f$University of California Irvine, Irvine, CA  92697, 
$^g$University of California Santa Cruz, Santa Cruz, CA  95064, 
$^h$Cornell University, Ithaca, NY  14853, 
$^i$University of Cyprus, Nicosia CY-1678, Cyprus, 
$^j$University College Dublin, Dublin 4, Ireland,
$^k$University of Edinburgh, Edinburgh EH9 3JZ, United Kingdom, 
$^l$University of Fukui, Fukui City, Fukui Prefecture, Japan 910-0017
$^m$Kinki University, Higashi-Osaka City, Japan 577-8502
$^n$Universidad Iberoamericana, Mexico D.F., Mexico,
$^o$Queen Mary, University of London, London, E1 4NS, England,
$^p$University of Manchester, Manchester M13 9PL, England, 
$^q$Nagasaki Institute of Applied Science, Nagasaki, Japan, 
$^r$University of Notre Dame, Notre Dame, IN 46556,
$^s$University de Oviedo, E-33007 Oviedo, Spain, 
$^t$Texas Tech University, Lubbock, TX  79609, 
$^u$IFIC(CSIC-Universitat de Valencia), 46071 Valencia, Spain,
$^v$University of Virginia, Charlottesville, VA  22904,
$^w$Bergische Universit\"at Wuppertal, 42097 Wuppertal, Germany,
$^{ee}$On leave from J.~Stefan Institute, Ljubljana, Slovenia, 
}}
\noaffiliation

\begin{abstract}
We present a measurement of the top quark mass with $\ttbar$ dilepton events 
\mbox{produced in $\ppbar$ collisions} at the Fermilab Tevatron ($\sqrt{s}$=1.96 TeV) and collected by the CDF II detector. 
A sample of 328 events with a charged electron or muon and an isolated track, corresponding to an integrated luminosity 
of $2.9$ fb$^{-1}$, are selected as $\ttbar$ candidates.
To account for the unconstrained  event  kinematics, we  scan over the phase space 
of the azimuthal angles ($\phi_{\nu_1},\phi_{\nu_2}$) of 
neutrinos and  reconstruct the top quark mass for each $\phi_{\nu_1},\phi_{\nu_2}$ pair  by minimizing a $\chi^2$ function  
in the $\ttbar$ dilepton  hypothesis. 
We assign  $\chi^2$-dependent weights  to
the solutions in order to build a preferred mass for each event.
Preferred mass distributions (templates) are built from simulated $\ttbar$ and background events, and parameterized in 
order to provide continuous probability density functions. 
A likelihood fit to the mass distribution in data as a weighted sum of signal and background 
probability density functions gives a top quark mass of 
$165.5^{ +{3.4}}_{-{3.3}}$(stat.)$\pm 3.1$(syst.)\ GeV/$c^2$.  
\end{abstract}

\pacs{14.65.Ha, 13.85.Qk, 12.15.Ff}

\maketitle

\section{\label{sec:Intro}Introduction}

The standard model (SM) 
explains the non-zero weak boson masses by spontaneous
breaking of the electroweak (EW) symmetry induced by the Higgs field~\cite{HiggsBoundaries}.
Also, non-zero quark masses are generated by the
coupling of the Higgs doublet with the fundamental fermions. However, their values
are not predicted since they are proportional to the unknown Yukawa couplings of each
quark.
The enormous top quark mass, which has a value comparable to the EW 
scale, justifies the  suspect 
that this quark may play a special role in the mechanism which breaks EW symmetry. 
In addition, because of 
its large mass, the top quark 
gives the largest contribution to loop corrections in the $W$~propagator. 
Within the SM, the  correlation between the top mass and the W mass 
induced by these corrections allows setting limits on the mass of the yet 
unobserved Higgs boson, and favor a relatively light Higgs. 
A more accurate measurement of the top quark mass will tighten the SM predicted 
region for the Higgs boson mass. 

According to the SM, at the Tevatron's 1.96 TeV 
energy top quarks are dominantly produced in pairs, by $\qqbar$ annihilation in $\sim$ 85$\%$ of the cases 
and by gluon fusion in the remaining $\sim$ 15$\%$~\cite{top_15_85}. 
Due to its extremely short lifetime, which in the SM  is expected to be about $10^{-25}$ s,  
the top quark decays before hadronizing
in $\sim$ 100$\%$ of cases into 
a $W$ boson and $b$-quark~\cite{PDG}.
Subsequently the $W$ boson can either decay into quarks as a $\qqbar '$ pair or into a charged lepton-neutrino pair. 
This allows classifying the $\ttbar$ candidate events into three final states: all-hadronic, lepton+jets, or dilepton, depending on the decay modes of the two $W$ bosons in the event. 
The all-hadronic state, where both $W$'s decay hadronically (about 46$\%$ of $\ttbar$ events), is characterized by six or more jets in the event. 
The lepton+jets final state contains one electron or muon (about 30$\%$ of $\ttbar$ events), four or more jets, and one neutrino.
Analyses dealing with the lepton+jets final state have provided the most precise top quark mass measurements, due to an optimal compromise between statistics and backgrounds. 
The dilepton final state, which is defined by the presence of two leptons (electrons or muons, about 5$\%$ of $\ttbar$ events), 
two or more jets, and large 
missing transverse energy 
from the two neutrinos, is the cleanest one, but suffers from the poorest statistics.\\
\indent It is important to perform measurements using independent data samples in all final states in order 
to improve the precision on the top quark mass and to be able to cross-check the results. Once the 
channel-specific SM backgrounds have been removed, discrepancies in the results across different samples 
could provide hints of new physics. The present analysis is performed in the dilepton final state 
by means of \mbox{lepton+track (``LTRK'')} top-pair selection. This selection is chosen to collect a large portion 
of events (about 45\%) not involved in the other CDF high precision top mass analyses performed 
in the dilepton final state~\cite{dil_neuro,dil_ljets}. \\
\indent The paper reports a measurement of the top quark mass
 with data collected by CDF II  before spring 2008, 
corresponding
to $2.9$ fb$^{-1}$ of integrated luminosity.
We select $\ttbar$ candidate  events in dilepton channel by
requiring a well identified electron
or muon plus a second, more loosely defined lepton, which is an isolated
track.
The measurement of the top quark mass in this channel is particularly challenging because of the two neutrinos in the final state. The kinematics is under-constrained and therefore assumptions on some missing final state observables are needed in order to reconstruct the event. 
In order to constrain the kinematics, 
we scan over the space of possibilities for the azimuthal angles of the two neutrinos,
and reconstruct the top quark mass by minimizing a $\chi^2$ function using the $\ttbar$ dilepton hypothesis.
A weighted average over a grid of the azimuthal neutrino angles   ($\phi_{\nu_1},\phi_{\nu_2}$) returns a single top quark mass value per event. 
In this analysis the Breit-Wigner probability distribution function with a top quark mass--dependent decay width
is applied 
in the kinematical event reconstruction,
which helps to decrease the statistical 
uncertainty by  20\% compared to the method  described in \cite{Previous Analysis Art}.
The top quark mass distribution in the data is fitted to the parameterized signal and background templates,
and the mass is extracted as the one corresponding to the best fit.\\ 
\indent Sections \ref{TheCDFDetector} and \ref{DataSample} describe the detector and the selection of the data sample. Section \ref{AnalysisOverview} gives an overview of the method used to reconstruct the events and to derive a single value of the top quark mass for each event. Section \ref{LikelihoodFit} defines the parameterization of signal and background mass distributions and the likelihood function used to fit the data
to these distributions.
 Section \ref{Calibration of the Method} describes the studies performed to calibrate the method, sections \ref{Results} and \ref{Systematics} 
 present the results and the systematic uncertainties, and
section \ref{conclusions} gives the conclusions.

\section{The CDF II Detector}
\label{TheCDFDetector}

The Collider Detector at Fermilab was upgraded in the year 2000 (CDF II, Figure~\ref{fig:cdfel}) in order to be able to handle the higher collision rate from the increased Tevatron luminosity. CDF II is a cylindrically and forward-backward symmetric apparatus detecting the products of the $\ppbar$ collisions over almost the full solid angle. A cylindrical ($r$, $\phi$, $z$) coordinate system is used to describe the detector geometry. The origin of the reference system  is the geometric center of the detector, with the $z$ axis pointing along the proton beam. 
The pseudorapidity $\eta$ is defined by $\eta \equiv -\ln (\tan (\theta/2))$, where $\theta$ is the polar angle relative to the $z$ axis.
The detector elements which are most relevant for this analysis are described below.
A more complete description of the detector can be found elsewhere~\cite{CDF}.

The tracking system consists of an inner silicon system and an outer gas drift chamber, 
the Central Outer Tracker (COT). 
The entire tracker is enclosed in a superconducting solenoid which generates a nearly uniform 1.4 T magnetic field in the $z$ direction and provides precision tracking and momentum measurement of charged particles within $|\eta|\leq 1$. The silicon tracker, which covers the $|\eta|<2$ region, is composed of the innermost detector (L00)~\cite{L00}, the Silicon Vertex Detector (SVXII)~\cite{SVXII}, and the Intermediate Silicon Layers (ISL)~\cite{ISL}. L00 is a layer of single-sided radiation-hardened silicon strips mounted directly on the beam pipe at a radius ranging from 1.35 cm to 1.62 cm. SVXII is an approximately 95 cm long cylinder of five layers of double-sided 
silicon microstrips 
covering a radial region between 2.5 cm and 10.7 cm. The ISL employs the same sensors as SVXII and covers the radial region between 20 cm and 28 cm, with one layer in the central region and two layers at larger angles. 
The COT~\cite{COT}, which spans 310 cm in length at a radial distance  ranging between 43 and 132 cm,
 contains four axial and four $\pm2 \degree$ stereo superlayers of azimuthal drift cells.
Axial and stereo superlayers alternate radially with one another.
The COT provides full coverage in the $|\eta|\leq 1$ region,
with reduced coverage in the region $1 < |\eta|\leq 2$.

Sampling calorimeters, divided into an inner electromagnetic and an outer hadronic compartment, surround the solenoid. 
Except for limited areas of non-instrumented regions (``cracks''), the calorimeters provide full azimuthal coverage within $|\eta|\leq 3.6$. 
All calorimeters are split into towers with projective geometry pointing at the nominal interaction vertex~\cite{CDF}. 
Embedded in the electromagnetic compartment, a shower maximum detector provides  good position measurements of the electromagnetic showers and is used
in electron identification~\cite{PES}.

The muon detection system consists of stacks of drift chamber modules backed by plastic scintillator counters. The stacks are four layer deep with laterally staggered cells from layer to layer to compensate for cell-edge inefficiencies. Four separate systems are used to detect muons in the $|\eta|<1.5$ region. The central muon detector (CMU)~\cite{CMU} is located behind the central hadronic calorimeter at a radius of $\sim$ 3.5 m from the beam axis, covering the $|\eta|<0.63$ region. The central muon upgrade detector (CMP) is arranged to enclose the $|\eta|<0.54$ region in an approximate four-sided box. It is separated from the CMU by the additional shielding provided by 60 cm of steel. The central muon extension (CMX) extends the muon identification to the region $0.6<|\eta|<1.0$. The more forward region ($1.0<|\eta|<1.5$) is covered by the intermediate muon detector (IMU). Table \ref{CDFdetector} summarizes  the characteristics of the CDF sub-detectors used in this  analysis.

\begin{ctable1c}
\caption{
CDF II sub-detectors, purposes, resolutions or acceptances.
}
\label{CDFdetector}
\begin{ruledtabular}
\begin{footnotesize}
\begin{tabular}{ccccp{0.2cm}}
Component				&Purpose		&Resolution/Acceptance		&Reference\\
\hline
\bf Silicon System				&Hit position		&11 $\mu m$ (L00)		&\cite{L00}\\ 
					&			&9 $\mu m$ (SVXII)		&\cite{SVXII}\\
					&			&16$\div$23 $\mu m$ (ISL)	&\cite{CDF}\\ 
  					&Impact parameter	&40 $\mu m$			&\\	
					&Interaction Vertex Position &70 $\mu m$		&\\	
\\
\bf COT					&Hit position		&140 $\mu m$			&\cite{COT}\\
					&Momentum measurement	&$\frac{\sigma_{P_T}}{P_T}=0.15\times P_T [GeV/c]$&\\
\\
\bf Central Calorimeters			&&\\
Electro-magnetic calorimeter		&Energy 	&$\frac{\sigma_{E}}{E}=13.5\%/\sqrt{E_T[GeV]}\oplus 2\%$&\cite{CEM}\\
Shower Max Detector			&Position 	&2 $mm$							&\cite{CES}\\
Hadron Calorimeter 			&Energy		&$\frac{\sigma_{E}}{E}=50.0\%/\sqrt{E[GeV]}\oplus 3\%$&\cite{CHA}\\
Wall Hadron Calorimeter 		&Energy		&$\frac{\sigma_{E}}{E}=75.0\%/\sqrt{E[GeV]}\oplus 4\%$&\cite{CDF}\\
\\
\bf Forward Calorimeters			&&\\
Electro-magnetic calorimeter		&Energy &$\frac{\sigma_{E}}{E}=16.0\%/\sqrt{E[GeV]}\oplus 1\%$	&\cite{PEM}\\
Shower Max Detector			&Position 	&1 $mm$							&\cite{PES}\\
Hadron Calorimeter			&Energy		&$\frac{\sigma_{E}}{E}=80.0\%/\sqrt{E[GeV]}\oplus 5\%$&\cite{CDF}\\
\\
\bf Muon Systems					&Muon Detection \\
CMU						&		&$P_T>$1.4 GeV/$c$				&\cite{CMU},\cite{CDF}\\
CMP						&		&$P_T>$2.2 GeV/$c$				&\cite{CDF}\\
CMX						&		&$P_T>$1.4 GeV/$c$				&\cite{CDF}\\
\end{tabular}
\end{footnotesize}
\end{ruledtabular}
\end{ctable1c}

CDF uses a three-level trigger system to select events to be recorded on tape, 
filtering the interactions from a 1.7 MHz
average bunch crossing rate to an output of 75-100 Hz.
 This analysis uses data from triggers based on leptons with high transverse momentum  $P_T$, as expected from the leptonically decaying $W$'s in the event. 
 The first two trigger levels perform
 limited reconstruction using dedicated hardware, which reconstructs tracks from the COT in the $r-\phi$ plane with a 
 transverse momentum resolution
 better than 2$\%\times P_T^2$ [GeV/$c$]~\cite{XFT}. The electron trigger requires a coincidence of a COT track with an electromagnetic cluster in the central
 calorimeter, while the muon trigger requires that a COT track points toward a set of hits in the muon chambers.
The third level is a software trigger which runs offline algorithms optimized for speed.
\begin{cfigure}
\includegraphics[width=\columnwidth]{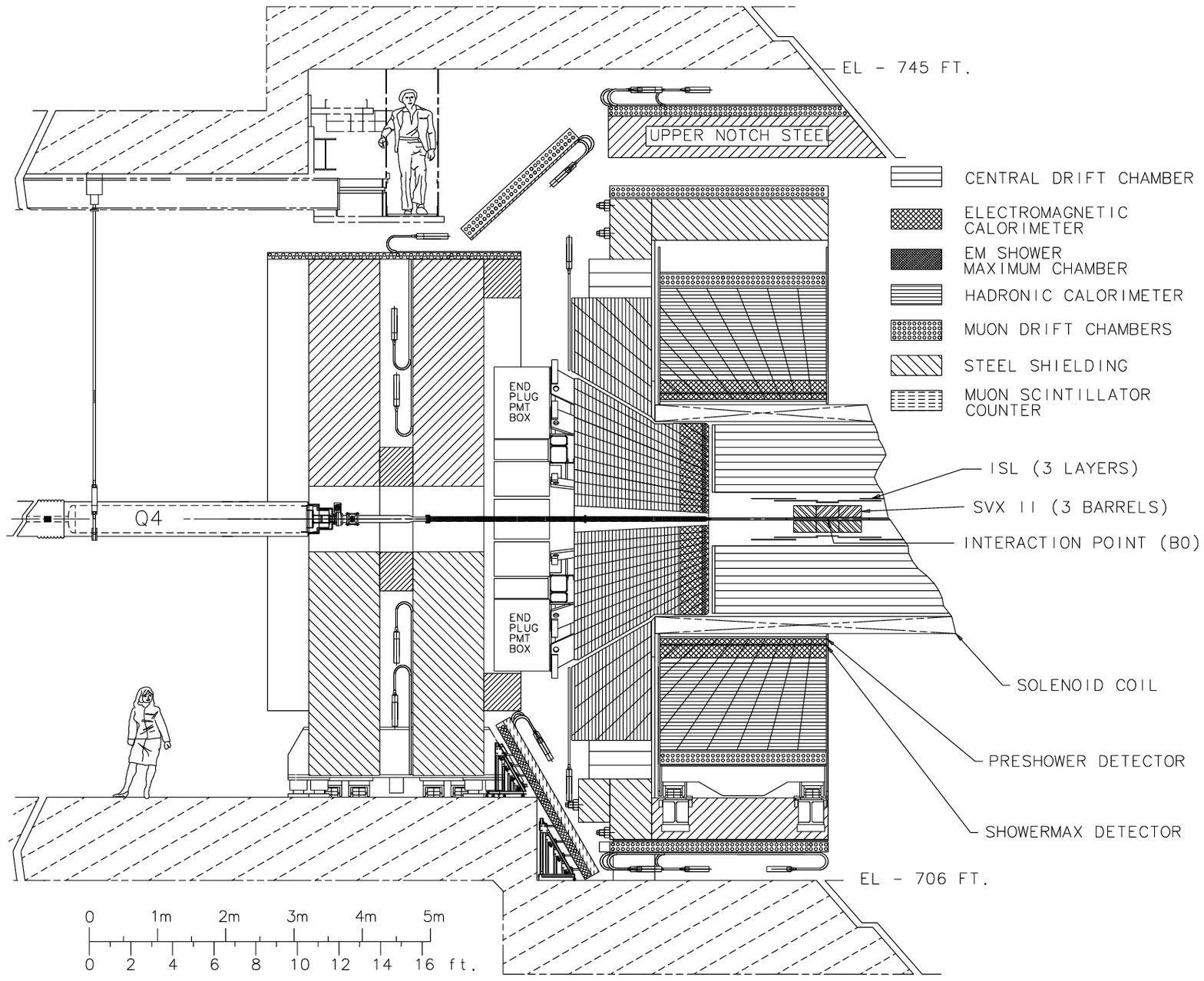}
\caption{Elevation view of half of the CDF II detector,
showing the inner microstrip detector, the Central Outer Tracker drift chamber, the
electromagnetic and hadronic calorimeters, the
muon drift chambers and scintillation counters.} \label{fig:cdfel}
\end{cfigure}

\section{Data Sample}
\label{DataSample}

The signature of $\ttbar$ dilepton events consists of two large transverse momentum leptons ($e$ or $\mu$), large missing transverse energy ($\MET$), two jets originating 
from $b$ quarks, and possible additional jets from initial and final state radiation. We select dilepton events from inclusive high-$P_T$ electron and muon triggers using the standard CDF lepton+track algorithm, as described in the next sections.\\ 
\indent The main expected background processes in the dilepton sample are $W$+jets with a jet misidentified as a lepton (``fakes''), Drell-Yan ($Z$/$\gamma^*\rightarrow e^+e^-,\mu^+ \mu^-,\tau^+ \tau^-$), and diboson events ($WW$, $WZ$, $ZZ$) with additional jets. 
In the case of Drell-Yan, \mbox{non-physical} $\MET$ can be faked by mis-measured jets or leptons. 
The contribution of these processes to the selected data sample is reduced by optimized selection cuts. 

\subsection{Trigger}
\label{Trigger}
A high-transverse-momentum lepton is required by the trigger. For a central electron candidate, an electromagnetic calorimeter cluster with $E_T \equiv E \cdot \sin\theta\geq18\ $GeV,  accompanied by a matched COT track with $P_T \equiv\ P \cdot \sin\theta\geq9\ $GeV/$c$, is required. For an electron in the plug region ($1.1<|\eta|<2.0$), the trigger requires an electromagnetic cluster in the calorimeter with $E_T\geq20\ $GeV and $\MET\geq15\ $GeV. For muon candidates two or more hits in the outer muon chambers matching a track of $P_T\geq18\ $GeV/$c$ in the central tracker are required.

\subsection{Leptons}
\label{Leptons}
The LTRK selection aims at selecting two charged leptons 
\mbox{of opposite} charge with a greater acceptance than if tight lepton selection cuts were applied on both leptons. One lepton (``tight lepton'') must have a well-measured track reconstructed from the interaction point with associated hits in the COT and SVX.
For muons, the track is required to be compatible with hits in the muon chambers and to have $P_T>20\ $GeV/$c$ and $|\eta|<1$. 
For forward electrons a calorimetry-seeded tracking algorithm is used to identify tracks since the plug region is not well covered by the COT.
In the case of electrons, the track is required to point to an electromagnetic cluster with $E_T>20\ $GeV and $|\eta|<2$.  Tight leptons must also satisfy an isolation requirement, i.e. the additional $E_T$ in a cone of radius $\Delta R=\sqrt{\Delta \eta^2+\Delta \phi^2}=0.4$ about the lepton trajectory must not exceed 10$\%$ of the lepton $E_T$. 

The other lepton (``track lepton'') is required to be a well-measured track originating at the interaction point with $|\eta|<1$ and $P_T>20\ $GeV/$c$. 
The track lepton must be isolated, which means that the ratio
between the additional transverse momentum of tracks in a $\Delta R=0.4$ cone around the track lepton 
and the overall $P_T$ in the cone is less than 10$\%$. Compared to the dilepton selection (``DIL''~\cite{DIL}) LTRK relaxes the calorimeter constraints on the track lepton in order to recover those events in which a lepton hits a detector crack. We refer to \cite{Previous Analysis Art} for a more detailed comparison between LTRK and DIL.

\subsection{Jets}
\label{Jets}
Jets are the final products of quark hadronization. They are identified by looking for clusters of energy in the calorimeter using the JETCLU cone algorithm~\cite{JETCLU}. The jet search is seeded by towers with $E_T>1\ $GeV. Starting from the most energetic seed, all seeds within
a 7$\times$7 bins wide area 
around the seed are grouped into a cluster and the centroid is calculated. Seeds cannot belong to more than one cluster. All towers with $E_T>0.1\ \textrm{GeV}$ within a $\Delta R=0.4$ cone about the cluster centroid are added to the cluster and the centroid is recalculated. The procedure is iterated and a final step of splitting and merging is performed in order not to include the same tower in more than one jet. 

Jet transverse energy is corrected for non-uniformities in the calorimeter response as a function of jet $\eta$, multiple $\ppbar$ interactions, and the hadronic jet energy scale of the calorimeter~\cite{JetCorr}. Events are required to have two or more jets with $E_T>20\ $GeV and $|\eta|<2$.

\subsection{Missing Transverse Energy}
\label{MissingTransverseEnergy}
The definition of the uncorrected missing transverse energy is:
\begin{equation}
\label{MET_def}
\overrightarrow{\MET}=-\sum_i E_T^i\hat{n}_i
\end{equation}
where the sum is performed over all towers with a deposited energy of at least 0.1 GeV.
$\hat{n}_i$ is the transverse unit vector pointing from the CDF geometrical center to the $i^{th}$ tower.

$\overrightarrow{{\MET}}$ is corrected to compensate for the following effects:
\begin{itemize}
 \item the interaction vertex displacement with respect to the CDF geometrical center: $\overrightarrow{\MET}$ is recalculated  with $\hat{n}_i$ (Eq. \ref{MET_def}) as having the origin in the interaction point.
 \item potential jet mis-measurement: if a track within the jet cone has a transverse momentum larger than the jet trasnverse energy, the difference between the $P_{T}$ of the highest-$P_T$ track and the jet $E_{T}$ is added to $\overrightarrow{\MET}$.
 \item muons: to correct  $\overrightarrow{\MET}$ for the  identified muons and to account for their minimum ionization contribution in the calorimeters, 
the difference between muon  calorimeter  $E_T$ and  muon  $P_T$ is added to  $\overrightarrow{\MET}$.  
 \item jet corrections: $\overrightarrow{\MET}$ is updated according to the corrections applied to the jet transverse energies, as explained above.
\end{itemize}

After corrections are applied the magnitude of the missing transverse energy is required to be larger than 25 GeV.

\subsection{Final Selection Cuts}
\label{FinalSelectionCuts}
Several topological vetoes are implemented in order to reduce the impact of backgrounds in the sample. Background contributions from $Z$ boson decays yielding overestimated $\MET$ are removed \mbox{by raising} the $\MET$ requirement to 40 GeV and
\mbox{the invariant}   mass of the tight lepton+track lepton pair
to be inside the $Z$ mass window ($[76,\ 106]\ $GeV/$c^2$). 
Large azimuthal separations between the $\overrightarrow{\MET}$ and jets ($\Delta \phi>25 \degree$), tight lepton ($\Delta \phi>5 \degree$), and track lepton ($\Delta \phi>5 \degree$, $\Delta \phi<175 \degree$) are required. These requirements have been implemented in order to reduce the number of events where mis-measured leptons or jets lead to overestimated $\MET$, mostly contributed by the Drell-Yan process. A lower cut on the  angle between the tight lepton and $\overrightarrow{\MET}$ is applied to reduce the acceptance for $Z/\gamma^* \rightarrow \mu\mu$ as electron+track, where high-$P_T$ muons are misidentified as electrons because of the emission of bremsstrahlung photons. The requirement of a minimum azimuthal angle between jets 
and $\overrightarrow{\MET}$ is dropped if $\MET>50\ $GeV, since such large values of missing transverse energy are not expected to arise from jet mis-measurements.\\
\indent Events with muons from cosmic rays or electrons originating from  the conversion of photons are removed. Cosmic muons are identified by requiring a delayed coincidence of the particle hits in the calorimeter~\cite{CosmicRayAndConversion}. 
Conversions are identified by pairing the electron track to an opposite sign track originating from a common vertex~\cite{CosmicRayAndConversion}.

\subsection{Sample Composition}
\label{SampleComposition}
Table \ref{2.XfbRates} summarizes the $\ttbar$ signal and background rates expected for a LTRK sample corresponding to an integrated luminosity of 2.9 fb$^{-1}$. 
Depending on the process, background rates are estimated using simulated or data events. Simulated events are generated with the \textsc{pythia}~\cite{Pythia} Monte Carlo program, which employs CTEQ5L~\cite{CTEQ} parton distribution functions, leading-order QCD matrix elements for the hard process simulation, and parton showering to simulate fragmentation and gluon radiation. A full simulation of the CDF II detector~\cite{CDFIISimulation} is applied. Diboson and $Z/ \gamma^* \rightarrow \tau^+\tau^-$ rates are estimated with simulated events, while $Z/ \gamma^* \rightarrow e^+e^-,\mu^+ \mu^-$ rates are estimated with a mixture of data and simulation. 
We use 
$Z/ \gamma^* \rightarrow e^+e^-,\mu^+ \mu^-$ simulated events to predict the ratio of events in different kinematic regions, while we use data to normalize the overall rates.
The expected fakes from $W$+jets and $\ttbar$ single lepton events with a jet misidentified as a lepton are estimated with $W$+jets data~\cite{CorinnePublic}. 
Signal acceptance and expected rate are evaluated using simulated $\ttbar$ events  with a cross-section of \mbox{6.7 pb}~\cite{ttbarCrossSection} and a top quark mass of 175 GeV/$c^2$. 

\begin{table}
\begin{center}
\caption{Expected numbers of $\ttbar$ signal and background
events with statistical uncertainties for the LTRK data sample.
A $\ttbar$ cross-section of 6.7 pb at a top quark mass of \mbox{175 GeV/$c^2$} is assumed.
}
\label{2.XfbRates}
\begin{ruledtabular}
\begin{tabular}{cc}
Process					&Expected number\\ 
\hline
Signal ($\ttbar$)				&162.6 $\pm$ 5.1\\
\hline
$WW$					&10.5 $\pm$ 1.0\\
$WZ$					&3.8 $\pm$ 0.3\\
$ZZ$					&0.9 $\pm$ 0.1\\
$Z/ \gamma^* \rightarrow e^+e^-$ 	&20.8 $\pm$ 6.0\\
$Z/ \gamma^* \rightarrow \mu^+\mu^-$	&9.1 $\pm$ 3.1\\
$Z/ \gamma^* \rightarrow \tau^+\tau^-$	&19.6 $\pm$ 2.4\\
Fakes					&80.2 $\pm$ 15.7\\
\hline
Total background				&145.0 $\pm$ 17.3\\
\end{tabular}
\end{ruledtabular}
\end{center}
\end{table}

\section{Mass Reconstruction}
\label{AnalysisOverview}
In this section we describe the procedure to reconstruct an event-by-event preferred top quark mass ($\MTreco$). In the next sections we will explain how the $\MTreco$ distribution is used to extract the top quark mass.

\subsection{Kinematics in the Dilepton Channel}
To reconstruct the $\ttbar$ event one needs to get 4-momenta for 
six final state particles, 24 values in total. These final state particles are two leptons and two neutrinos from $W$'s decays, 
as well as two jets originated from  the top-decay $b$ quarks.
Out of the 24 final quantities, 16 (jet and lepton 4-momenta) are measured, two ($\overrightarrow{{\MET}}$ components in the transverse plane) are obtained by assuming overall transverse momentum conservation, and five constraints are imposed on the involved particle masses 
($m_{W^-}=m_{W^+}=m_W$, where $m_W=80.4\ $GeV/$c^2$~\cite{PDG},
 $m_t=m_{\overline{t}}$, $m_\nu=m_{\overline{\nu}}=0$). 
The event kinematics is therefore under-constrained. One must
assume that at least one more parameter is known
in order to reconstruct the kinematics and solve for the top quark mass.

\subsection{Neutrino $\phi$ Weighting Method}
The method implemented in this work for reconstructing the top quark mass event by event is called  ``Neutrino $\phi$ Weighting Method''. This method was previously described in \cite{Previous Analysis Art}. 
In order to constrain the kinematics a scan over the space of possibilities for the azimuthal angles of the neutrinos ($\phi_{\nu_1},\phi_{\nu_2}$) is used.
A top quark mass is reconstructed by minimizing a chi-squared function ($\chi^2$) in the dilepton $\ttbar$ event hypothesis. 
The $\chi^2$ has two terms:
\begin{eqnarray}
\label{chi2}
\chi ^2\ = \chi_{reso}^2 + \chi_{constr}^2
\end{eqnarray}

The first term takes into account the detector uncertainties, whereas the second one constrains the parameters to the known physical quantities 
within their uncertainties. 
The first term is as follows:

\begin{eqnarray}
\begin{split}
\label{chi2reso}
\chi_{reso} ^2  & = \sum_{l=1}^2 \frac{(P_T^l - \tilde{P_T^l})^2}{{\sigma_{P_T}^l}^2} - 2\sum_{j=1}^2 \ln(\mathscr{P}_{tf}(\tilde{P_T^j}|P_T^j)) \\
 & + \sum_{i=x,y} \frac{(E_{U}^{i} - \tilde{E_{U}^{i}})^2}{{\sigma_{E_U}}^2} 
\end{split}
\end{eqnarray}

With the use of the tilde ($\sim$) we specify the parameters of the minimization procedure,
whereas variables without tilde 
represent the measured values.
$\mathscr{P}_{tf}$ are the transfer functions between $b$ quark and jets: they express the probability of measuring a jet transverse momentum $P_T^j$ from a $b$ quark with transverse momentum $\tilde{P_T^j}$. We will comment on $\mathscr{P}_{tf}$ in Section \ref{Transfer Functions}.
The sum in the first term is over the two leptons in the event;
the second sum loops over the two highest-$E_T$ (leading) jets, which are assumed to originate from the $b$ quarks (this assumption is true in about 70$\%$ of simulated $\ttbar$ events~\cite{Previous Analysis Art}). 

The third sum in Eq.~\ref{chi2reso} runs over the transverse components of the unclustered energy ($E_{U}^{x}$, $E_{U}^{y}$), which is defined as the sum of the energy vectors from the towers not associated with leptons or any leading jets. It also includes possible additional 
jets with $E_T>8\ $GeV within $|\eta|<2$. 

The uncertainties ($\sigma_{P_T}$) on the tight lepton $P_T$ used for identified electrons ($e$) and muons ($\mu$) 
are calculated as ~\cite{Previous Analysis Art}:
\begin{eqnarray}
\frac{\sigma_{P_T}^e}{P_T^e} \ &=&\ \sqrt{\frac{0.135^2}{P_T^e\textrm{[GeV/$c$]}}\ +\ 0.02^2 }\\
\frac{\sigma_{P_T}^{\mu}}{P_T^{\mu}} \ &=&\ 0.0011\cdot P_T^{\mu}\textrm{[GeV/$c$]}
\end{eqnarray}
The track-lepton momentum uncertainty is calculated as for the muons, since momentum is measured in the tracker for both electrons and muons.
Uncertainty for the transverse components of the unclustered
energy, $\sigma_{E_U}$, is defined as
$0.4\sqrt{{E_T^{{\rm uncl}}}}$[GeV]~\cite{bjetsSyst}, where $E_T^{{\rm uncl}}$ is the scalar sum of the transverse energy 
excluding the two leptons and the two leading jets.

The second term in Eq.~\ref{chi2}, $\chi_{constr}^2$, constrains 
the parameters of the minimization procedure 
through the invariant masses of the lepton-neutrino and of the lepton-neutrino--leading-jet systems.
This term is as follows:

\begin{eqnarray}
\label{chi2constr}
\begin{split} 
\chi_{constr} ^2 = & -2\ln(\mathscr{P}_{BW}(m_{inv}^{l_1, \nu_1}|m_W, \Gamma_{m_W})) \\
&  -2\ln(\mathscr{P}_{BW}(m_{inv}^{l_2, \nu_2}|m_W, \Gamma_{m_W}))\\
& - 2\ln(\mathscr{P}_{BW}(m_{inv}^{l_1, \nu_1, j_1}|\tilde{m_t}, \Gamma_{\tilde{m_t}})) \\
& - 2\ln(\mathscr{P}_{BW}(m_{inv}^{l_2, \nu_2, j_2}|\tilde{m_t}, \Gamma_{\tilde{m_t}}))
\end{split}
\end{eqnarray}

$\tilde{m_t}$ is the parameter giving the reconstructed top quark mass. $\mathscr{P}_{BW}(m_{inv};\ m,\Gamma) \equiv \frac{\Gamma^2\cdot m^2}{(m_{inv}^2\ -\ m^2)^2\ +\ m^2\Gamma^2}$  indicates the relativistic Breit-Wigner distribution function, which expresses the probability that an unstable particle of mass $m$ and decay width $\Gamma$ decays into a system of particles with invariant mass $m_{inv}$. 
We use the PDG~\cite{PDG} values for $m_W$ and $\Gamma_{m_W}$.
For the top width we use the function 
\begin{equation}
\Gamma_{m_t}=\frac{G_F}{8\sqrt{2}\pi}m_t^3(1-\frac{m_W^2}{m_t^2})^2(1+2\frac{m_W^2}{m_t^2})
\label{Gamma Mt equation}
\end{equation}
according to Ref.~\cite{standard_model}. This  new formulation of the  $\chi_{constr} ^2$ term helps to decrease  the  statistical error of the top mass
reconstruction by 20\%.

The longitudinal components of the neutrino momenta are free parameters of the minimization procedure, while the transverse components are related to $\overrightarrow{\MET}$ and to the assumed $(\phi_{\nu_1},\phi_{\nu_2})$ as follows:
\begin{equation}
\label{PHI nutrino solutions}
\left\{
\begin{array}{cl}
P_x^{\nu_1} \equiv &  P_T^{\nu_1} \cdot \cos(\phi_{\nu_1})= \\
& \frac{{\met_T}_x \cdot \sin(\phi_{\nu_2}) -{\met_T}_y \cdot \cos(\phi_{\nu_2})}{\sin(\phi_{\nu_2}-\phi_{\nu_1})}
\cdot \cos(\phi_{\nu_1})  \\[.18in]
P_y^{\nu_1} \equiv &P_T^{\nu_1} \cdot \sin(\phi_{\nu_1})= \\
&\frac{{\met_T}_x \cdot \sin(\phi_{\nu_2}) -{\met_T}_y \cdot \cos(\phi_{\nu_2})}{\sin(\phi_{\nu_2}-\phi_{\nu_1})}
\cdot \sin(\phi_{\nu_1})  \\[.18in]
P_x^{\nu_2} \equiv &P_T^{\nu_2} \cdot \cos(\phi_{\nu_2})= \\
&\frac{{\met_T}_x \cdot \sin(\phi_{\nu_1}) -{\met_T}_y \cdot \cos(\phi_{\nu_1})}{\sin(\phi_{\nu_1}-\phi_{\nu_2})}
\cdot \cos(\phi_{\nu_2})  \\[.18in]
P_y^{\nu_2} \equiv & P_T^{\nu_2} \cdot \sin(\phi_{\nu_2})= \\
&\frac{{\met_T}_x \cdot \sin(\phi_{\nu_1}) -{\met_T}_y \cdot \cos(\phi_{\nu_1})}{\sin(\phi_{\nu_1}-\phi_{\nu_2})}
\cdot \sin(\phi_{\nu_2}) 
\end{array}
\right.
\end{equation}

The minimization procedure described above must be performed for all the allowed values of $\phi_{\nu_1}$, $\phi_{\nu_2}$ in the $(0,2\pi)\times(0,2\pi)$ region. 
Based on simulation, we choose 
a  $\phi_{\nu_1},\ \phi_{\nu_2}$ grid of 24$\times$24 values  as inputs for the minimization procedure. 
In building the grid we avoid the singular points at 
$\phi_{\nu_1}=\phi_{\nu_2}+k\cdot\pi$, where $k$ is integer. For these  points, which correspond to a configuration where  the  
two  neutrinos are collinear  in transverse plane,  the  kinematics of the  event  cannot be reconstructed using  Eqs.~\ref{chi2reso}-~\ref{PHI nutrino solutions}. 
Avoiding  these points  in  our procedure  does not effect  the  reconstruction  of the  top  mass  central  value,  
but rather effects the width of the  mass  distribution per event. 
Note from Eq.~\ref{PHI nutrino solutions} that performing the transformation 
$\phi_{\nu} \rightarrow \phi_{\nu}+\pi$
leaves $P_x^{\nu}$ and $P_y^{\nu}$ unchanged, but reverses the sign of $P_T^{\nu}$.
We exclude unphysical solutions ($P_T^{\nu_1} < 0$ and/or $P_T^{\nu_2} < 0$) and  choose the solution which leads to positive transverse momenta for both neutrinos. 
This decreases the number of grid points to 12$\times$12. 
At each point 8 solutions can exist, because of the two-fold ambiguity in the longitudinal momentum for each neutrino and of the ambiguity on the lepton-jet association. Therefore, for each event, we perform 1152 minimizations, each of which returns a value of $m^{reco}_{ijk}$ and $\chi^2_{ijk}$ ($i,j=1,\dots,12;\ k=1,\dots,8$). 
We define $\chi_{ij}'^2=\chi_{ij}^2 + 4\cdot \ln(\Gamma_{m_t})$,
which is obtained by using Eq. \ref{chi2constr} where $\mathscr{P}_{BW}$ is substituted with
$\frac{\Gamma\cdot m^2}{(m_{inv}^2\ -\ m^2)^2\ +\ m^2\Gamma^2}$,
 and select the lowest $\chi'^2$ solution for each point of the $(\phi_{\nu_1},\phi_{\nu_2})$ grid, thereby reducing the number of obtained masses to 144. Each mass is next weighted according to
\begin{equation}
\label{weight1}
w_{ij}\ =\ \frac{e^{-\chi_{ij}'^2/2}}{\sum_{i=1}^{12}\sum_{j=1}^{12}e^{-\chi_{ij}'^2/2}}
\end{equation}
A top quark mass distribution is built in order to identify 
the most probable value (MPV) for the event.
Based on a result of the simulation, the following procedure for improving
the performance of solution-weighting was implemented.
Masses below a threshold of 30$\%$ the MPV bin content are discarded, and the remaining ones are averaged to compute the preferred top quark mass for the event.

\subsection{Transfer Functions}
\label{Transfer Functions}
Since jet energy corrections have been calibrated on samples dominated by light quarks and gluons, we need an additional correction for a better reconstruction of the energy of $b$-quark jets. In Equation \ref{chi2reso}, we introduced the transfer functions $\mathscr{P}_{tf}$, which allow us to step back from jets to partons. These functions of jet $\eta$ and $P_T$ are defined as the parameterization of $\xi \equiv (P_T^{b-quark}-P_T^{jet})/P_T^{jet}$ distributions, built from a large sample of simulated $\ttbar$ events.
The $b$-quark jets in the simulation are recognized  using true MC information.
Jets with axis within a $R=0.4$ cone about the generated $b$ quarks are used.
The influence of $b$-quark $P_T$ spectra on the $\xi$ distributions is  minimized  
by choosing the weights inversely proportional to the probability density of $P_T^{b-quark}$.
Also, this 
greatly reduces
dependence of the transfer functions on $m_t$.
 
In order to parameterize the above distributions we found the following expression to be adequate:
\begin{eqnarray}\label{tf}
\begin{split} 
\mathscr{W}_{TF}(\xi) &=
\frac{\gamma_7\gamma_6}{\sqrt{2\pi}\gamma_2}\ e^{-0.5(\frac{\xi-\gamma_1}{\gamma_2}
+ \exp(-\frac{\xi -\gamma_1}{\gamma_2}))} \\
&+\ \frac{\gamma_7(1-\gamma_6)}{\sqrt{2\pi}\gamma_5}\ e^{-0.5(\frac{\xi
-\gamma_4}{\gamma_5})^2} \\
&+\ \frac{(1-\gamma_7)}{\sqrt{2\pi}\gamma_3}\ e^{-0.5(\frac{\xi
-\gamma_8}{\gamma_3})^2}
\end{split}
\end{eqnarray}

The parameters $\gamma_1 \cdots \gamma_8$ are derived from  the fit.
The distributions are built for three $|\eta|$ regions: $|\eta| < 0.7$, $0.7 < |\eta| < 1.3$, and $1.3 < |\eta| < 2.0$.
 
Figure~\ref{Top Specific Corrections} shows the distributions and 
the transfer functions for a number of $(|\eta|,P_T^{jet})$ regions. 10 GeV/$c$ wide $P_T$ bins are used from 30 GeV/$c$  to 190 GeV/$c$ 
for $|\eta| < 0.7$, from 30 GeV/$c$ to 150 GeV/$c$ for $0.7 < |\eta| < 1.3$, and from 30 GeV/$c$ to 110 GeV/$c$ for $1.3 < |\eta| < 2.0$. A single bin is 
used above  and below these  regions.

\begin{cfigure1c}
\epsfig{file=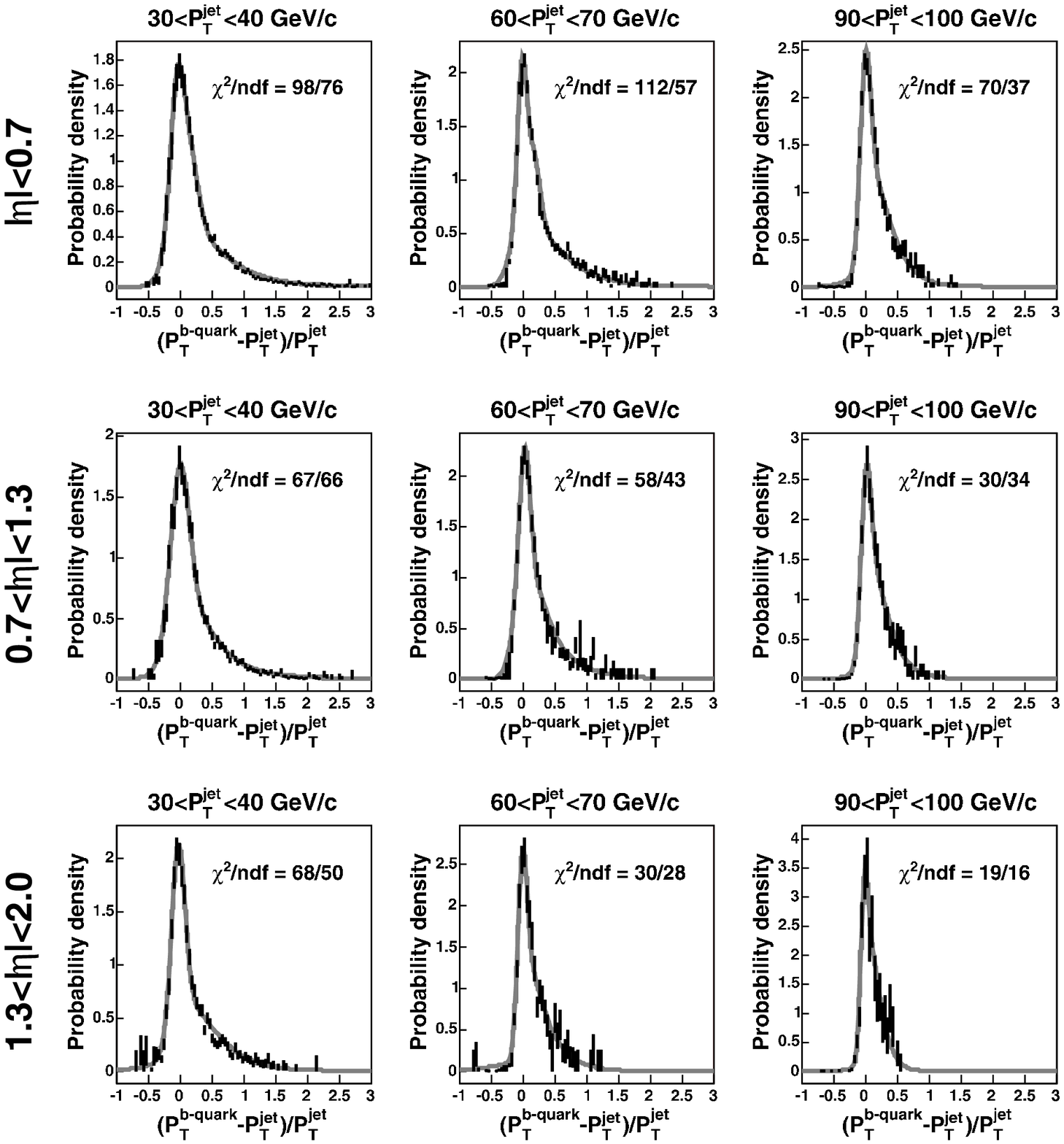 ,width=\textwidth,height=5.5in} 
\caption{Examples of the transfer functions of $b$ quarks into jets used in the fit. 
These functions of jet $\eta$ and $P_T$ are defined as the parameterization of $(P_T^{b-quark}-P_T^{jet})/P_T^{jet}$ distributions.
The points are from the simulated $\ttbar$ events. The curves show the parameterization
with Eq. \ref{tf}.
}
\label{Top Specific Corrections}
\end{cfigure1c}

\section{Top Quark Mass Determination}
\label {LikelihoodFit}
The selected data sample is a mixture of signal and background events. In order to extract the top quark mass, the reconstructed top quark mass distribution in data is compared with probability density functions (p.d.f.'s) for signal and background by means of a likelihood minimization. P.d.f.'s are defined as the parameterizations of $\MTreco$ 
templates obtained by applying the neutrino $\phi$ weighting method on simulated signal and background events, which are selected according to the lepton+track algorithm.

\subsection{Templates}
\label{Templates}
Signal templates are built from $\ttbar$ samples generated with \textsc{pythia} for 
top quark masses in the range $155$ to $195$ GeV/$c^2$ in 2~GeV/$c^2$ steps. 
They are parameterized in a global fit by using a combination of one Landau and two Gaussian distribution functions, as:
\begin{eqnarray}
\label{Signal Template parameterization}
\begin{split} 
P_s(\MTreco|m_t) &= \frac{c_1 p_6}{\sqrt{2\pi}p_2}\ e^{-0.5(\frac{\MTreco-p_1}{p_2} + \exp(-\frac{\MTreco -p_1}{p_2}))} \\
&+ \frac{c_1(1-p_6)}{\sqrt{2\pi}p_5}\ e^{-0.5(\frac{\MTreco -p_4}{p_5})^2} + \\
&\frac{(1-c_1)}{\sqrt{2\pi}p_3}\ e^{-0.5(\frac{\MTreco -c_2}{p_3})^2}
\end{split} 
\end{eqnarray}
$P_s$, the signal p.d.f., expresses the probability that a mass $m_t^{reco}$ is reconstructed from an event with true top quark mass $m_t$. 
The constants $c_1$ and $c_2$ are set a-priori to adhere to 
the features of the template shape.
The parameters $p_1,\dots,p_6$  depend on the true top quark mass $m_t$ are calculated as: 
\begin{equation}
\label{pk_from_alpha}
p_k = \alpha_k + \alpha_{k+6} \cdot (m_{t}\textrm{[GeV/$c^2$]}-175)\quad k=1,\dots,6
\end{equation}
The parameters $\alpha_k$ are obtained from the fit to the signal templates.
Figure \ref{Signal Templates} shows a subset of templates along with their parameterizations (solid lines).
\begin{cfigure1c}
\epsfig{file=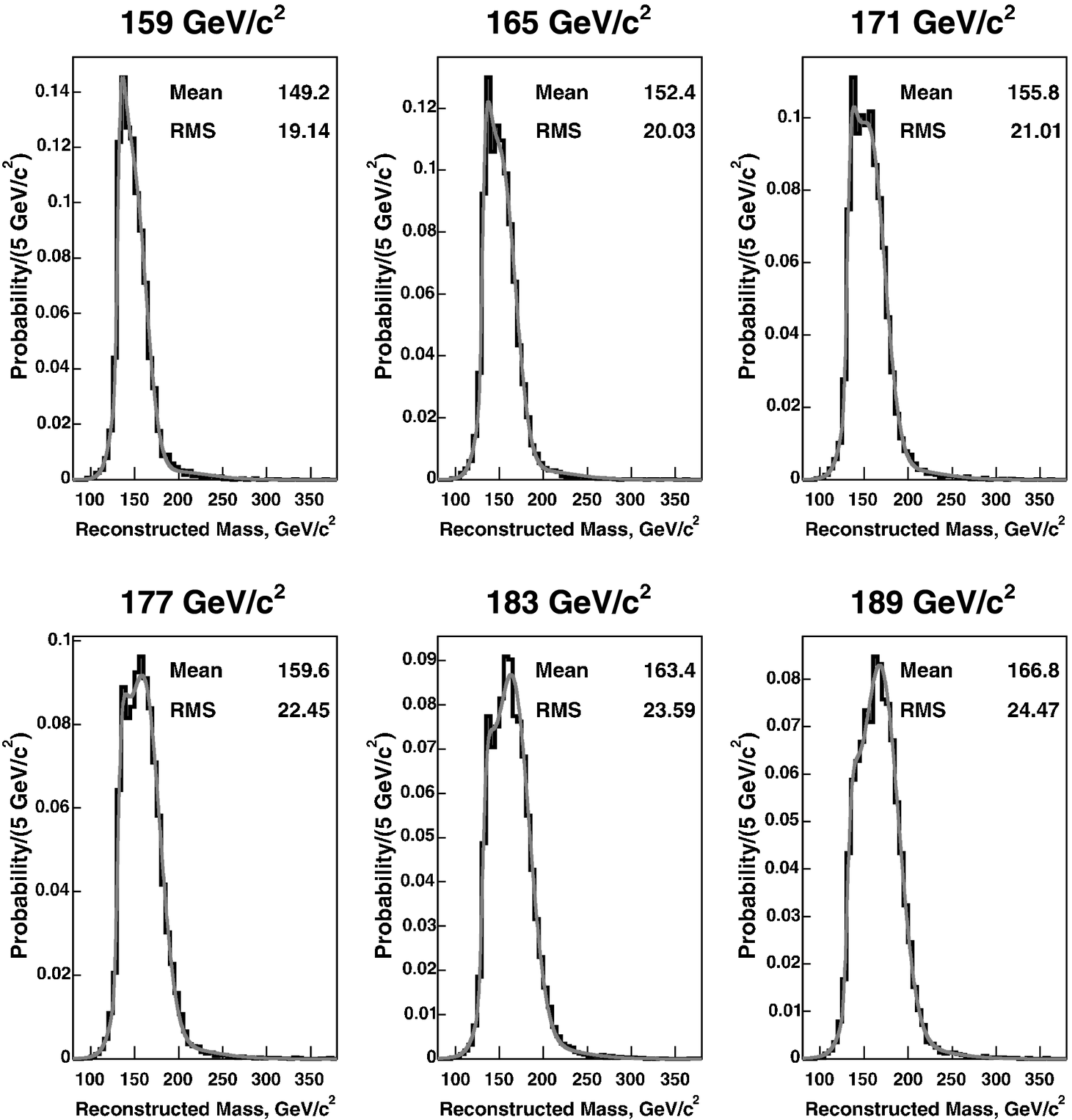,width=0.8\textwidth,height=4.5in} 
\caption{
Signal templates and fitting functions (solid lines) for a number of generated top quark masses. The parameterization is defined in Eq. \ref{Signal Template parameterization}.
}
\label{Signal Templates}
\end{cfigure1c}

A representative background template is built by adding fakes, Drell-Yan, and diboson  templates. These templates have been normalized to the expected rates reported in Table \ref{2.XfbRates}. The fakes template is built from $W$+jets data events by weighting each event 
according to the probability for a jet to be mis-identified as a lepton (fake rate)~\cite{CorinnePublic}. Drell-Yan and diboson templates are built from samples simulated 
with \textsc{pythia} and \textsc{alpgen}~\cite{ALPGEN} + \textsc{pythia} respectively. 
The combined background template is fitted with a sum of two Landau and one Gaussian distribution functions, as:

\begin{eqnarray}
\label{Background Template parameterization}
\begin{split} 
P_b(\MTreco) &= \frac{k_1\beta_6}{\sqrt{2\pi}\beta_2}\ e^{-0.5(\frac{\MTreco-\beta_1}{\beta_2}+\exp(-\frac{\MTreco -\beta_1}{\beta_2}))}\\
&+\frac{k_1(1-\beta_6)}{\sqrt{2\pi}\beta_5}\ e^{-0.5(\frac{\MTreco-\beta_4}{\beta_5})^2}\\ 
&+\frac{(1-k_1)}{\sqrt{2\pi}\beta_3}\ e^{-0.5(\frac{\MTreco-k_2}{\beta_3}+ \exp(-\frac{\MTreco -k_2}{\beta_3}))}
\end{split}
\end{eqnarray} 
where the fit parameters $\beta_1\cdots \beta_6$ are $m_t$-independent. 
The constants $k_1$ and $k_2$ are set a-priori to adhere to the features  
of the template shape. 
The combined background template and its parameterization (solid line), Drell-Yan, diboson, and fakes templates are plotted in Figure \ref{Background Templates}. 
\begin{cfigure1c}
\epsfig{file=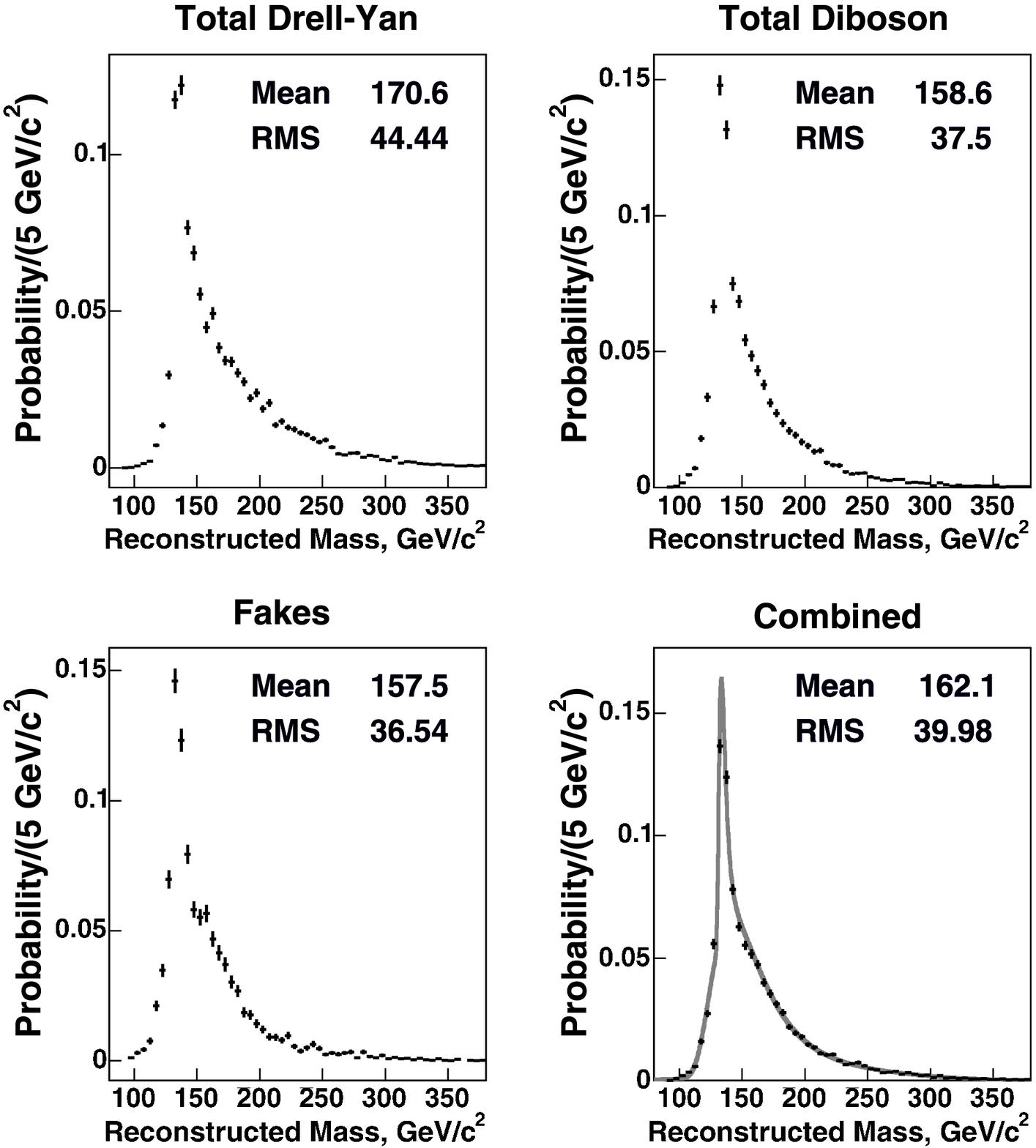,width=0.6\textwidth}
\caption{
Drell-Yan, fakes, diboson and combined background templates. The fitting function (solid line), defined in Eq. \ref{Background Template parameterization}, is superimposed to the combined template.
}
\label{Background Templates}
\end{cfigure1c}

\subsection{Likelihood Minimization}
\label{LikelihoodMinimization}
The top quark mass estimator is extracted from the data sample by performing an unbinned likelihood fit and minimization. The likelihood function expresses the probability that a $\MTreco$ distribution from data is described by a mixture of background events and dilepton $\ttbar$ events with an assumed top quark mass. 
Inputs for the likelihood fit are the reconstructed mass ($m_n$), the simulated signal and background p.d.f.'s, and the expected background. The
background expectation ($n_b^{exp}$=145.0) and its 
uncertainty ($\sigma_{n^{exp}_b}$=17.3) are taken from Table~\ref{2.XfbRates}.
The likelihood takes the form:
\begin{eqnarray}
\label{General likelihood}
\mathscr{L}= \mathscr{L}_{shape} \cdot \mathscr{L}_{backgr} \cdot \mathscr{L}_{param};
\end{eqnarray}
where
\begin{eqnarray}
\label{Lshape}
\begin{split}
\mathscr{L}_{shape}& =
\frac{e^{-(n_s+n_b)} \cdot (n_s+n_b)^N}{N!} \\ 
&\cdot
\prod_{n=1}^{N}
\frac{n_{s} 
\cdot P_{s}(m_{n}|m_{top})+n_b \cdot P_{b}(m_{n})}{n_s+n_b},
\end{split}
\end{eqnarray}
\begin{equation}
\label{nb likelihood}
\mathscr{L}_{backgr}=
\exp(\frac{-(n_b-{n_b^{exp}})^2}{2 \sigma^2_{n^{exp}_b}})
\end{equation}
and
\begin{equation}
\begin{split}
\mathscr{L}_{param} &= \exp\{
-0.5[(\vec{\alpha}-\vec{\alpha_0})^TU^{-1}(\vec{\alpha}-\vec{\alpha_0}) \\ 
&+(\vec{\beta}-\vec{\beta_0)}^TV^{-1}(\vec{\beta}-\vec{\beta_0})]\}.
\label{Lpar}
\end{split}
\end{equation}
The top quark mass estimator ($m_{top}$) returned by the minimization is the mass corresponding to $[-\ln\mathscr{L}]_{min}$. 
The shape likelihood term, $\mathscr{L}_{shape}$ (Eq.~\ref{Lshape}), expresses the probability of an event being signal with the top mass $m_{top}$ or background. The signal ($P_s$)
and background ($P_b$) probabilities are weighted according to the number of signal ($n_s$) and background ($n_b$) events, which are floated in the likelihood fit. In the fitting procedure, $n_b$ is constrained to be Gaussian-distributed with mean value $n_b^{exp}$ and standard deviation $\sigma_{n^{exp}_b}$, as shown in Eq. \ref{nb likelihood}, while $(n_s+n_b)$ is the mean of a Poisson distribution of N selected events. In this manner, the number of signal events is independent of the expected $\ttbar$ lepton+track events in a particular assumption of the $\ttbar$ cross-section value.
$\mathscr{L}_{param}$ constrains the parameters of the signal ($\vec{\alpha}$) (see Eq.~\ref{pk_from_alpha}) and background ($\vec{\beta}$) (see Eq.~\ref{Background Template parameterization}) p.d.f.'s.
These p.d.f.'s have Gaussian distribution with mean values ($\vec{\alpha}_0$) and ($\vec{\beta}_0$) obtained from the signal and background templates fit.
$U$ and $V$ are the corresponding covariant matrices for $\vec{\alpha}$ and $\vec{\beta}$ 
returned from the MINUIT~\cite{MINUIT} minimization.
\section{Calibration of the Method}
\label{Calibration of the Method}
The method described above is calibrated in order to avoid systematic biases in the measured top quark mass and in its 
uncertainty. Calibrations are performed by running a large number ($10^4$) of ``pseudo-experiments'' (PE's) on simulated background and signal events where the true top quark mass is known. Each PE consists of determining the number of signal ($N^{PE}_s$) and background ($N^{PE}_b$) events in the sample, drawing $N^{PE}_s$ masses from a signal template and $N^{PE}_b$ from the background template, and fitting the mass distribution to a combination of  signal and background p.d.f.'s, as described in Section \ref{LikelihoodFit}. 
A top quark mass ($m_t^{fit}$) and its positive and negative statistical 
uncertainties ($\sigma^+$ and $\sigma^-$) are returned by the fit.
Numbers of signal and background events are generated according to Poisson distributions with means given in Table \ref{2.XfbRates}. 

For each input top quark mass the median of the $m_t^{fit}$ distribution is chosen as the  mass estimate ($m_t^{out}$). 
The distributions of $m_t^{out}$ versus input mass ($m_t$) and the bias, defined as $\Delta M=m_t^{out} - m_t$, are shown in  Figure~\ref{Sanity}.
The 
uncertainty bars are determined by the limited statistics of the signal and background templates.
\begin{cfigure}
\epsfig{file=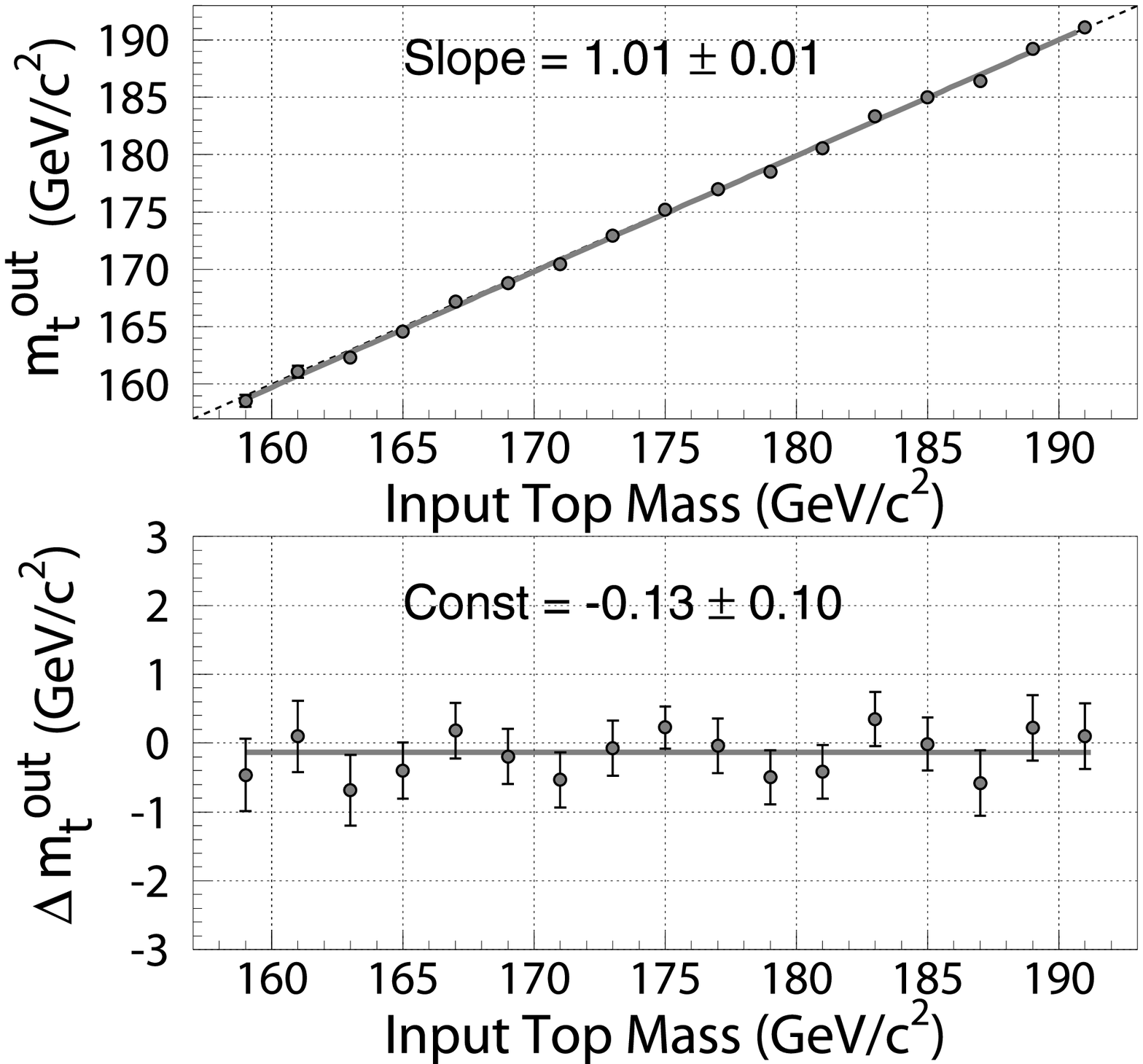,width=\columnwidth} 
\caption{
Results from pseudo-experiments. The upper plot shows $m_t^{out}$ versus input masses, while the lower one shows the bias.
}
\label{Sanity}
\end{cfigure}
Both fits in Figure~\ref{Sanity} are performed in the mass range 159-191 GeV/$c^2$. The slope of the straight line in the upper plot is consistent with one, while the average bias (horizontal line in the lower plot) is $-0.13 \pm 0.10\ $GeV/$c^2$. Although this value can be considered compatible with zero within uncertainties, 
we apply a shift of +0.13~GeV/$c^2$ to the result on data.

In order to check the bias on the statistical 
uncertainty we use pull distributions,
defined as follows:
\begin{equation}
\label{Alternative pull definition}
\mathrm{pull}=\frac{m_t^{fit}\ -\ m_t}{\sigma'}
\end{equation}
where $\sigma'=\big\{{\sigma^+\ \textrm{if}\ m_t^{fit}\ <\ m_t\atop{|\sigma^-|\ \textrm{if}\ m_t^{fit}\ >\ m_t}}$.
The  positive and negative statistical 
uncertainties are returned by MINUIT (routine MINOS)~\cite{MINUIT}. 
For each generated top quark mass, pull distributions are fitted by Gaussian functions (some examples are shown in Figure~\ref{Pulls example}).
\begin{cfigure}
\epsfig{file=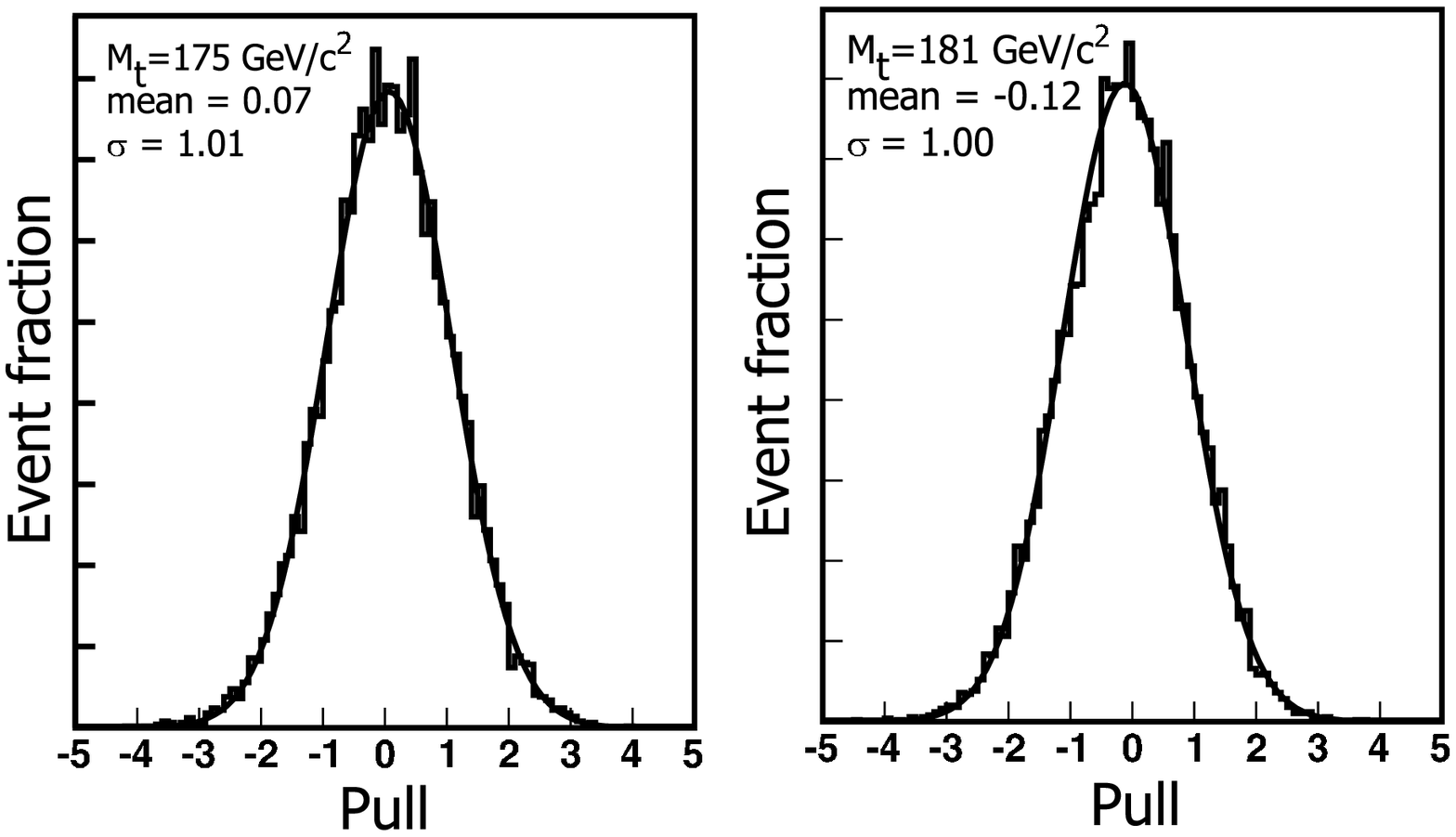,width=\columnwidth}
\caption{
Results from pseudo-experiments: pull distributions for generated mass samples at $m_t = 175\ $GeV/$c^2$ (left) and $m_t = 181\ $GeV/$c^2$ (right). 
Distributions are fitted to Gaussian functions (solid line), returning the indicated means and standard deviations.
}
\label{Pulls example}
\end{cfigure}\\
\indent The mean and width of the pull distributions versus generated top quark mass are shown in Figure~\ref{Mean and Width of the pull}. Error bars account for the limited statistics of signal and background templates.
\begin{cfigure}
\epsfig{file=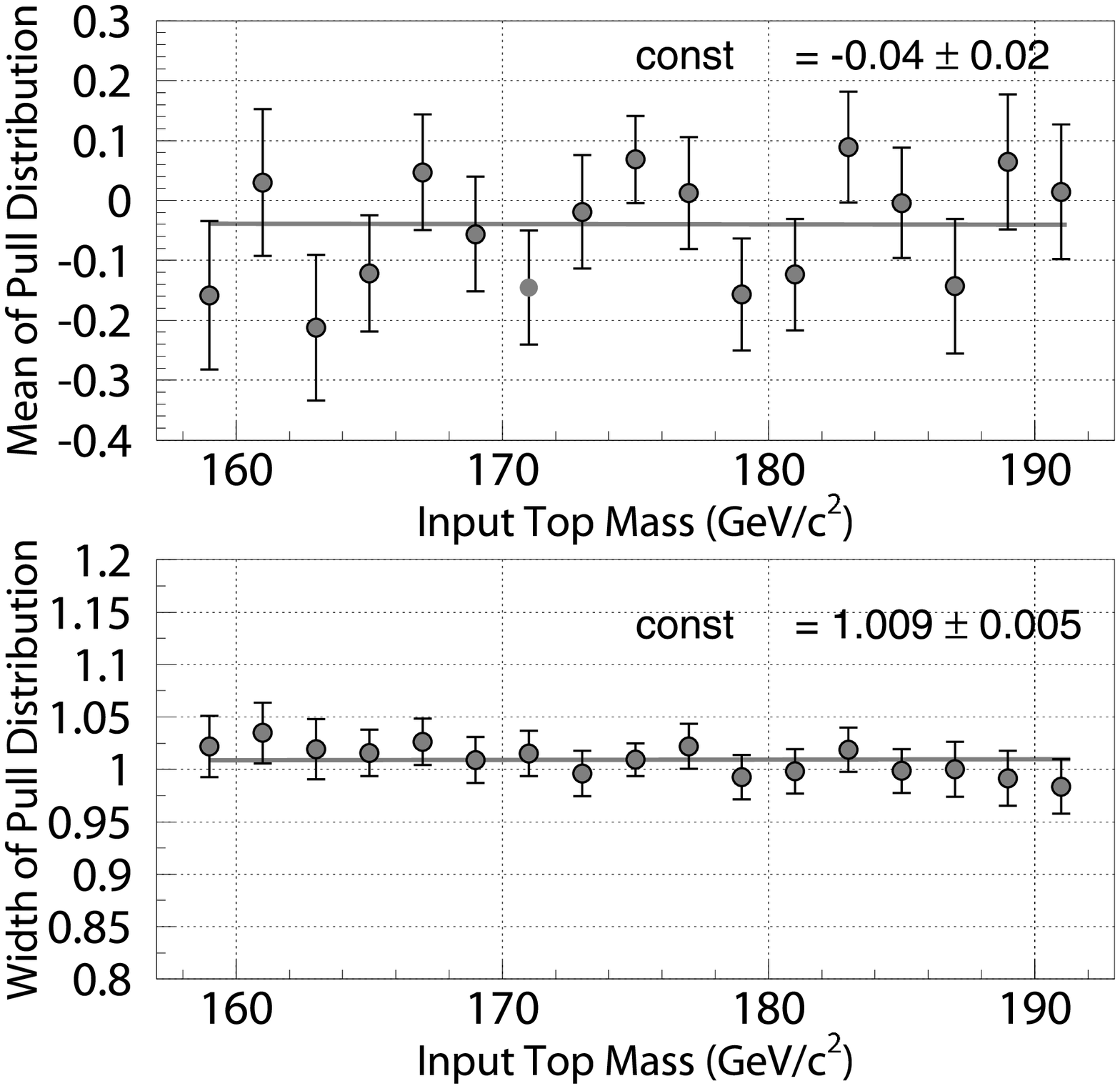,width=\columnwidth}
\caption{
Results from pseudo-experiments: mean and width of the pull distributions versus generated top quark mass are shown in the upper and lower plots respectively.
}
\label{Mean and Width of the pull}
\end{cfigure}
The average width of pull distributions is $1.009 \pm 0.005$. A  width larger than one indicates an underestimate of the statistical 
uncertainty. Accordingly, the statistical 
uncertainty obtained from data is increased  by a factor 1.009. 

\section{Results}
\label{Results}
The data sample used in this measurement corresponds to an integrated luminosity of  $2.9$ fb$^{-1}$. 
A total of 328 LTRK candidates are found in data. Selected events are reconstructed and an experimental mass distribution is built. The likelihood constrained fit described in Section \ref{LikelihoodMinimization} is performed 
and the following estimate of the top quark mass with statistical 
uncertainties is obtained:
\begin{equation}
m_{top}\ =\ 165.35^{+3.35}_{-3.22}\ \mbox{GeV/c$^2$}
\label{Top Mass Stat Error}
\end{equation}
The experimental top quark mass distribution is shown in Figure~\ref{Constrained Likelihood Fit}. 
\begin{cfigure}
\epsfig{file=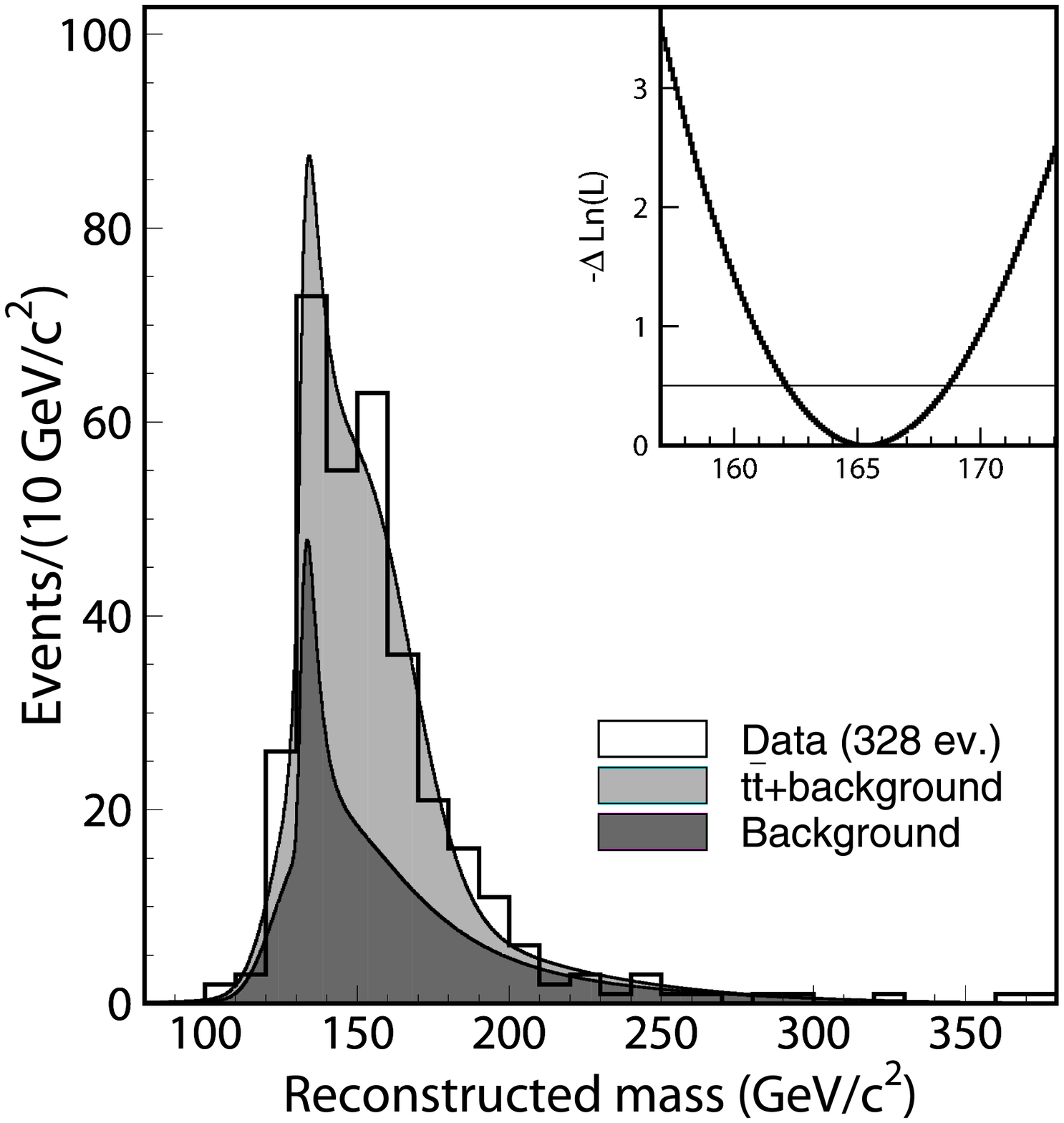,width=\columnwidth}
\caption{Two-component constrained fit to the 328-event LTRK data sample. Background (dark gray) and signal+background (light gray) p.d.f.'s, normalized according to the numbers returned by the fit, are superimposed to the reconstructed mass distribution from data (histogram). The insert shows the fitted mass-dependent negative log-likelihood function.
}
\label{Constrained Likelihood Fit}
\end{cfigure}
The constrained fit returns 
$181.4^{+21.9}_{-21.3}$ signal and $146.1^{+15.1}_{-15.0}$ background events.
\mbox{The observed} rates are in good agreement with expectations (Table \ref{2.XfbRates}).\\
\indent As a check, we remove the Gaussian constraint on the number of background events in Formula \ref{General likelihood}. The unconstrained fit returns:
 \begin{equation}
m_{top}\ =\ 165.33^{+3.39}_{-3.28}\ \mbox{GeV/$c^2$}
\label{Unconstrained Top Mass Stat Error}
\end{equation}
\noindent with $178.6^{+30.9}_{-31.1}$ signal and $149.4^{+31.6}_{-29.5}$ background events. 
The top quark mass and the number of signal and background events from unconstrained and constrained fits are in agreement. \\
\indent The top quark mass and its statistical 
uncertainty obtained from the constrained fit (Eq.~\ref{Top Mass Stat Error})
are corrected
  for the expected systematic 0.13~GeV/$c^2$ shift, 
and for the 1.009 width of the pull distribution (Section \ref{Calibration of the Method}) respectively. The final value is:
\begin{eqnarray}
m_{top}\ =\ 165.5^{+3.4}_{-3.3} \mathrm{(stat.)}\ \mbox{GeV/$c^2$}
\label{Top Mass Stat Error Final}
\end{eqnarray}
In order to check that the measured statistical 
uncertainty is reasonable, a set of PE's is performed on simulated background and signal events with $m_t=165\ $GeV/$c^2$ (close to  the central value of the constrained fit), as explained in Section \ref{Calibration of the Method}. The obtained positive and negative error distributions along with the observed values (arrows) are shown in Figure~\ref{Expected Vs data errors}.
\begin{cfigure}
\epsfig{file=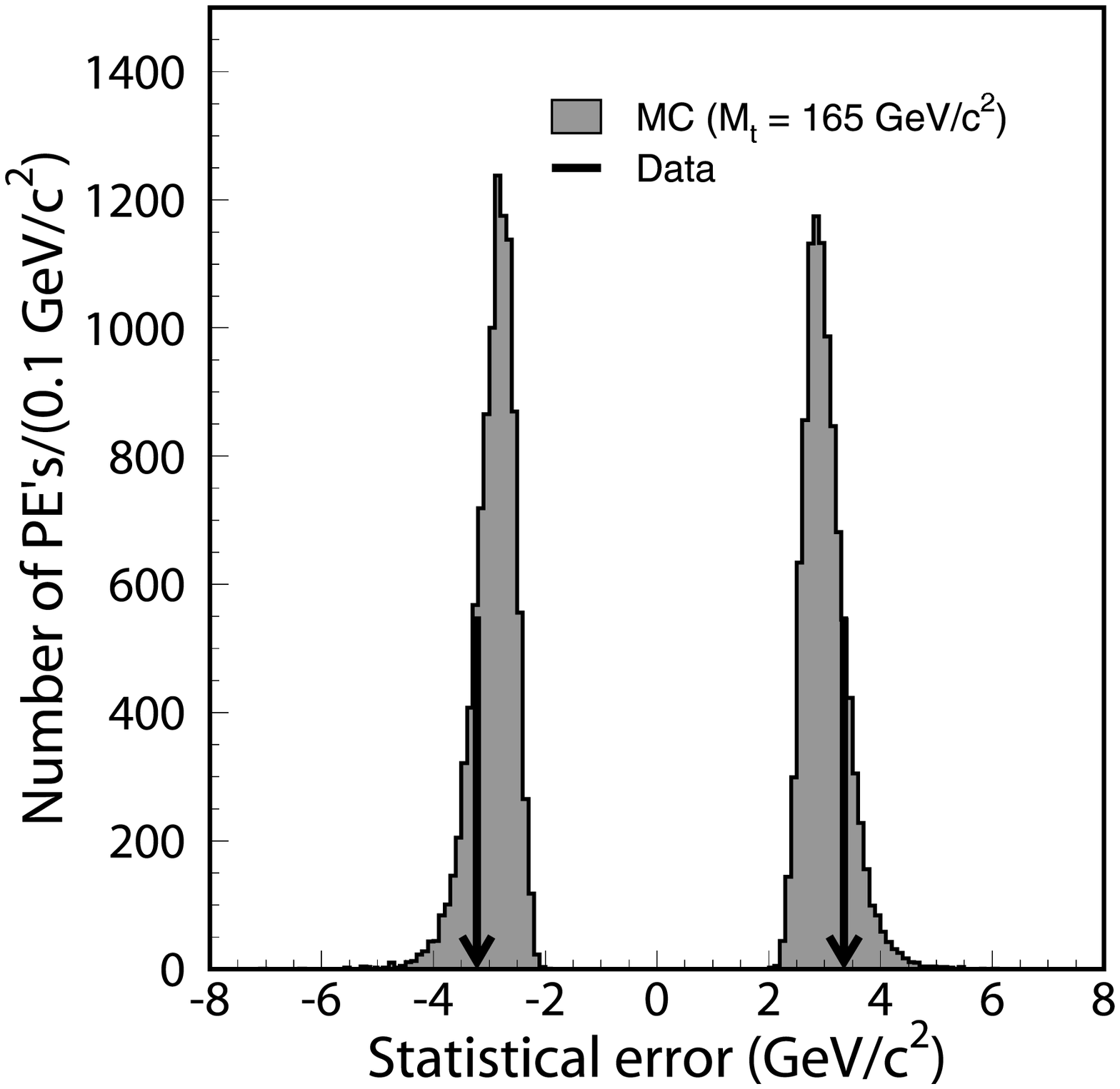,width=\columnwidth}
\caption{
Expected statistical 
uncertainties from pseudo-experiments generated with a top quark mass of 165~GeV/$c^2$. The arrows indicate the 
uncertainties found in this measurement.}
\label{Expected Vs data errors}
\end{cfigure}
We found that the probability for obtaining 
a precision
better than that found in this experiment is  $82\%$.

\section{Systematic Uncertainties}
\label{Systematics}
Since our method compares findings to expectations estimated from Monte Carlo simulations, uncertainties in the models used to generate events cause systematic uncertainties. Other systematic uncertainties arise from the potential mis-modeling of the background template shape.

The procedure for estimating a systematic uncertainty is as follows. The parameters used for the generation of events are 
modified by $\pm$ 1 standard deviation in their uncertainties and new templates are built. PE's from the modified templates are performed using the same p.d.f.'s as in the analysis. 
The obtained medians of the top quark mass distribution from PE's and the nominal top quark mass are used to estimate the systematic uncertainty. 
The source of each systematic uncertainty is assumed to be uncorrelated to the other ones, so that the overall systematic 
uncertainty is obtained by adding in quadrature the individual uncertainties. The systematic uncertainties along with the total uncertainty are summarized in Table \ref{systerr_total}.
In the following, we describe how each systematic uncertainty is evaluated.

\subsection{Jet energy scale}
\label{Jet Energy Scale and b-jet Energy Scale}
The measured jet energy is corrected according to the measured and simulated calorimeter response to electrons and hadrons~\cite{Jet Corrections}. Jet corrections also correct for the non-uniformities in calorimeter response as a function of $|\eta|$, effects of multiple $\ppbar$ collisions, the hadronic jet energy scale, deposited energy within the jet cone by the underlying events, and \mbox{out-of}-cone jet energy lost in the clustering procedure. The systematic uncertainty due to the jet energy scale (JES) is estimated from signal and background events in which each jet energy correction has been shifted by $\pm$ 1 standard deviation in the energy scale factor. 
Shifted signal and background templates are built and two sets of $10^4$ PE's are performed. 
The systematic uncertainty for each level of corrections is taken as $(m_t^+-m_t^-)/2$, where $m_t^+$ and $m_t^-$ are the top quark masses found respectively for a lower and upper shift of the parameter. The individual uncertainties are summed in quadrature in order to obtain the JES systematic uncertainty. Results are reported in Table \ref{syst_energyscale}.
\begin{table}[htbp]
\begin{center}
\caption{
Summary of jet energy scale systematic uncertainties 
}
\begin{tabular}{lc}
\hline
\hline
Source 		& Uncertainty (GeV/$c^2$) \\
\hline
$\eta$ calorimeter non-uniformity & 0.6\\
multiple interactions & 0.0\\
hadronic jet energy scale 	& 2.2\\
underlying event  	&  0.2\\
out-of-cone energy loss		& 1.8\\
\hline
Total & 2.9\\
\hline
\hline
\end{tabular} 
\label{syst_energyscale}
\end{center}
\end{table}
The systematic uncertainty in the top quark mass due to  the JES uncertainty is 2.9 GeV/$c^2$. \\

Since jet energy corrections are estimated with studies dominated by light quarks and gluon jets, additional uncertainty occurs on the $b$-jet energy scale because of three main reasons~\cite{bjetsSyst}: 
\begin{enumerate}
 \item uncertainty in the heavy-flavor fragmentation model;
\item uncertainty in the $b$-jet semileptonic branching ratio; 
\item uncertainty in the calorimeter response to energy released by $b$-jets. 
\end{enumerate}
The effect of the fragmentation model on the top quark mass is evaluated by reweighting events
according to two different fragmentation models from fits on LEP~\cite{LEP} and SLD~\cite{SLD} data,
while effects of the uncertainties on the semileptonic $b$-jet branching ratio (BR) and $b$-jet energy calorimeter response 
are estimated by shifting the BR 
and the $b$-jet energy scale.
In all cases shifted templates are built and PE's are
performed. The resulting shifted masses are used to estimate the systematic uncertainty
due to each of the sources. These uncertainties are added in quadrature.
 The total systematic uncertainty in the $b$-jet energy scale is 0.4 GeV/$c^2$.

\subsection{Lepton energy scale}
\label{Lepton energy scale}
The uncertainty on the lepton energy scale 
may affect  the top quark mass measurement. This uncertainty is studied by applying a $\pm1\%$ shift to the $P_T$ of leptons~\cite{JetCorr}. 
Shifted templates are built and PE's are performed. Half of the difference of the resulting masses is taken as the systematic 
uncertainty on the top quark mass due to  the lepton energy scale uncertainty.
The systematic uncertainty in the lepton energy scale is 0.3 GeV/$c^2$.

\subsection{Monte Carlo event generation}
\label{Monte Carlo event generation}
Several systematic uncertainties are due to the modeling of $\ttbar$ signal events.

\subsubsection{ Monte Carlo generators}
The effect of the choice of a particular Monte Carlo generator is studied by comparing our default \textsc{pythia} generator to \textsc{herwig}.
These generators differ in the hadronization models, 
handling of the underlying $\ppbar$ events and of the multiple $\ppbar$ collisions in the same bunch crossing, 
and in the spin correlations in the production and decay of $\ttbar$ pairs (implemented in \textsc{herwig} only)~\cite{Pythia Herwig generator}. The difference between masses obtained from sets of PE's performed with the two generators is found. 
The  systematic uncertainty due to our choice of Monte Carlo generators is 0.2 GeV/$c^2$.

\subsubsection{Initial and final state radiation}
The effect of the initial and final state radiation (ISR and FSR) parameterization is studied, since jets radiated by interacting partons can be misidentified as leading jets and affect the top quark mass measurement. The systematic uncertainty associated with ISR is obtained by adjusting the QCD parameters in the DGLAP~\cite{DGLAP I} parton shower evolution in $\ttbar$ events. The size of this adjustment has been obtained from comparisons between Drell-Yan data and simulated events~\cite{bjetsSyst}. Since the physical laws that rule ISR and FSR are the same, the parameters that control ISR and FSR  are varied together (IFSR). Half of the difference in top quark mass from PE's performed on samples with increased and decreased IFSR is taken as the systematic uncertainty for the radiation  modeling. 
The systematic uncertainty due to uncertainties in the initial and final state radiation is 0.2 GeV/$c^2$.

\subsubsection{PDFs}
The uncertainty in reconstructing the top quark mass due to the use of sets of 
parton distribution function (PDF) comes from three sources: PDF choice, PDF parameterization, and QCD scale ($\Lambda_{QCD}$). The uncertainty due to the PDF choice is estimated as the difference between the top quark mass extracted by using CTEQ5L (default) and MRST72~\cite{MRST}. The uncertainty due to PDF parameterization is estimated by shifting by $\pm$1 standard deviation one at a time the 20 eigenvectors of CTEQ6M~\cite{CTEQ}. Half of the differences between the shifted masses derived from PE's are added in quadrature. The measured mass differences between MRST72, generated with $\Lambda_{QCD}=300$ MeV, and MRST75, generated with $\Lambda_{QCD}=228$ MeV, \cite{MRST} is taken as the uncertainty due to the choice of  $\Lambda_{QCD}$. These systematic uncertainties are added in quadrature. Results are summarized in Table \ref{PDF_syst}. 
 The total systematic uncertainty due to uncertainties in  the PDFs is 0.3 GeV/$c^2$.
\begin{table}[htbp]
\begin{center}
\caption{
PDF systematic uncertainties on top quark mass. The total systematic uncertainty is the sum in quadrature of the individual contributions.}
\begin{tabular}{lc}
\hline
\hline
Source   &          Uncertainty (GeV/$c^2$)  \\
\hline	 
PDF parameterization                    & 0.2 \\
PDF choice              & 0.1  \\
$\Lambda_{QCD}$     & 0.2 \\
\hline   
Total & 0.3 \\
\hline
\hline
\end{tabular}
\label{PDF_syst}
\end{center}
\end{table}

\subsubsection{Luminosity profile (event pileup)}
Pseudo-experiment simulations have only been made for a probability of multiple interactions in a single bunch crossing as appropriate for the collider luminosity during the first period of data taking ($1.2$ fb$^{-1}$ integrated luminosity.) A possible discrepancy between simulation and data collected at later times at higher luminosity may affect
the top quark mass measurement. We evaluate this effect by running batches of PE's on
$\ttbar$ events, selected according to the number of interaction vertices found in the event.

\begin{cfigure}
\epsfig{file=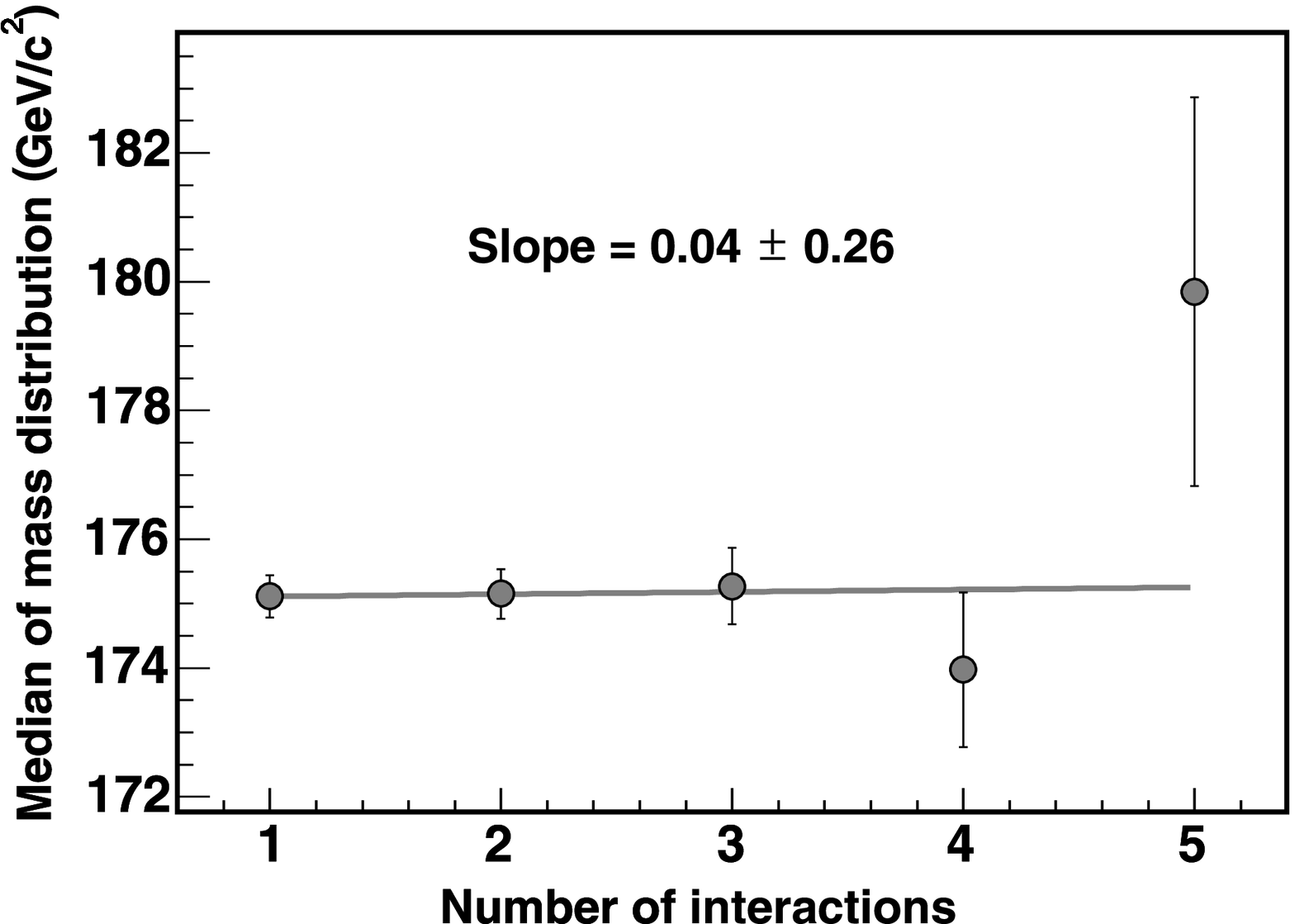,width=\columnwidth}
\caption{
Results from pseudo-experiments performed using events selected according to the number
of interactions.
}
\label{pileup}
\end{cfigure}
The results from PE's are plotted against the number of interactions and a linear fit is applied (Figure \ref{pileup}).
Since we do not see a significant mass dependence, we
use the uncertainty (0.26 GeV/$c^2$/interaction) on the slope to derive the systematic
uncertainty. We multiply 0.26 GeV/$c^2$/interaction by $<N^{data}_{vtx}>-<N^{MC}_{vtx}>$, 
where $<N^{data}_{vtx}>=$2.07 and  $<N^{MC}_{vtx}>=$1.50 are the average number of vertices in the selected
data sample and simulated sample respectively. 
 We obtain a 0.15 GeV/$c^2$ top mass uncertainty
due to the event pile-up.

\subsection{Background template shape}
\label{Background template shape}
The systematic uncertainties due to the potential mis-modeling of the background template shape were also estimated. We identify three independent sources for this systematic uncertainty: background composition, $W$+jets fakes shape, and Drell-Yan shape. The effect of the diboson shape is neglected because of the small expected rate of this background (Table \ref{2.XfbRates}).

In order to estimate the systematic uncertainty for the background composition, fakes, diboson, and Drell-Yan, the expected rates are alternatively varied by plus or minus one standard deviation (Table \ref{2.XfbRates}) without changing the total number of expected background events. Half of the differences between $\pm$ 1 $\sigma$ shifted masses derived from PE's are added in quadrature. 
 The systematic uncertainty due to uncertainty in the background composition is 0.5 GeV/$c^2$. 

The uncertainty on the shape of the fake background template (Section \ref{Templates}) is modeled. The fake rate $E_T$ dependence is varied 
according to the fake rate uncertainties in each $E_T$ bin. Two shifted background templates are built and  used for PE's. 
The corresponding shift in mass is taken as the systematic uncertainty due to potential mis-modeling of fake shape. 
 The top mass uncertainty due to uncertainty in the fake shape is 0.4 GeV/$c^2$. 

Drell-Yan events with associated jets can pass the selection because jet mis-measurements can cause a large unphysical $\MET$. Mis-modeling of this effect is studied, since it may affect the top quark mass measurement. Two modified Drell-Yan templates are built by re-weighting $Z/ \gamma^* \rightarrow e^+e^-, \mu^+ \mu^-$ events. The weight has been optimized by looking at discrepancies in $\MET$ between Monte Carlo simulation and data. Results of PE's performed with the modified Drell-Yan templates are used to estimate the systematic uncertainty due to the possible fluctuation in the shape of this background. 
 The mass systematic uncertainty due to uncertainties in the shape of the Drell-Yan background is 0.3 GeV/$c^2$.

\begin{table}[htbp]
\begin{center}
\caption{Summary of systematic uncertainties on the top quark measurement. 
}
\begin{tabular}{lc}
\hline
\hline
 Source & Uncertainty (GeV/$c^2$) \\
\cline{1-2}
Jet energy scale  & 2.9 \\
$b$-jet energy scale  & 0.4   \\
Lepton energy scale    &    0.3   \\
Monte Carlo generators & 0.2 \\
Initial and final state radiation & 0.2 \\
Parton distribution functions & 0.3  \\
Luminosity profile (pileup) & 0.2  \\
Background composition & 0.5 \\
Fakes shape  & 0.4         \\
Drell-Yan shape     & 0.3      \\ 
\cline{1-2} 
{\bf Total}  & {\bf 3.1} \\
\hline
\hline
\end{tabular}
\label{systerr_total}
\end{center}
\end{table}

\section{Conclusions}
\label{conclusions}
Using the template technique on a lepton+track sample we measure a top quark mass of 
\begin{equation}
\begin{array}{c}
m_{top}\ =\ 165.5^{+{3.4}}_{-{3.3}}\textrm{(stat.)}\pm 3.1\textrm{(syst.)}\ \mbox{GeV}/c^2 \\
  \textrm{or}    \\
m_{top}\ =\ 165.5^{+{4.6}}_{-{4.5}}\mbox{GeV}/c^2 \\
\end{array}
\end{equation}

This result agrees with the world average top quark mass ($m_{top} = 172.4 \pm 1.2\ $GeV/$c^2$~\cite{TopCombination}), obtained by combining the main CDF and D\O\ Run I (1992-1996) and Run II (2001-present) results.

Compared with our previous result ($m_{top} = 169.7 \pm 9.8 $ GeV/$c^2$~\cite{Previous Analysis Art} ), 
obtained 
on a $\int Ldt=340\ \mbox{pb}^{-1}$ data sample, 
a significant improvement in the total 
uncertainty has been achieved. The improvement due to the novelties in the analysis technique is estimated from PE's to be about 20$\%$. 
The improvements which made this progress possible are the introduction of relativistic Breit-Wigner
distribution functions in event reconstruction, along with $m_{top}$-dependent top width, 
while in \cite{Previous Analysis Art} Gaussian distribution functions and a constant top width were used. 
A new feature of this analysis is the use of a larger statistics lepton+track sample which overlaps by only $\sim$ 45\% with the
often used dilepton sample~\cite{Previous Analysis Art}.

\section{Acknowledgments}
We thank the Fermilab staff and the technical staffs of the participating institutions for their vital contributions. This work was supported by the U.S. Department of Energy and National Science Foundation; the Italian Istituto Nazionale di Fisica Nucleare; the Ministry of Education, Culture, Sports, Science and Technology of Japan; the Natural Sciences and Engineering Research Council of Canada; the National Science Council of the Republic of China; the Swiss National Science Foundation; the A.P. Sloan Foundation; the Bundesministerium f\"ur Bildung und Forschung, Germany; the Korean Science and Engineering Foundation and the Korean Research Foundation; the Science and Technology Facilities Council and the Royal Society, UK; the Institut National de Physique Nucleaire et Physique des Particules/CNRS; the Russian Foundation for Basic Research; the Ministerio de Ciencia e Innovaci\'{o}n, and Programa Consolider-Ingenio 2010, Spain; the Slovak R\&D Agency; and the Academy of Finland.

\newpage
%
%
%
%
%
%

%

\end{document}